\documentclass[twocolumn]{svjour3}
\usepackage{amssymb}
\usepackage{amsmath}
\usepackage{graphicx}
\usepackage{amsfonts}
\providecommand{\U}[1]{\protect\rule{.1in}{.1in}}

\begin{document}

% the only way i found to deal with multiple institutes

\title{Neural Control of Redundant (Abundant) Systems as Algorithms Stabilizing Subspaces}
\author{V. M. Akulin$^{1,2,3}$, and F. Carlier$^{1}$ \\
and \\
Stanislaw Solnik$^{4,5}$, and M.L. Latash$^{4,6}$ \\ \\
\normalfont{
$^{1}$Laboratoire Aim\'{e} Cotton, CNRS II, B\^{a}timent 505, 91405 Orsay Cedex, France \\
$^{2}$Laboratoire Jean-Victor Poncelet, CNRS, 11 Bolshoy Vlasyevskiy Pereulok, Moscow,119002, Russia \\
$^{3}$Institute for Information Transmission Problems of the Russian Academy of Science, Bolshoy Karetny per. 19, Moscow, 127994, Russia \\
$^{4}$Pennsylvania State University, University Park PA 16802, USA \\
$^{5}$University School of Physical Education, Wroclaw, Poland \\
$^{6}$Moscow Institute of Physics and Technology,9 Institutskiy per., Dolgoprudny, Moscow Region, 141700, Russia 
}}

\maketitle

% The abstract is after the maketitle. If not, it appears alone on the first page after compilation (?).

\begin{abstract}
We address the problem of stability of motor actions implemented by the
central nervous system based on simple algorithms potentially reflecting
physical (including physiological) processes within the body. A number of 
conceptually simple algorithms that solve motor tasks with a high probability
 of success may be based on feedback schemes that ensure stability of 
 subspaces of neural variables associated with accomplishing those tasks.
 The task is formulated in terms of linear
constrains imposed either on the human body mechanical variables or on neural
variables; we discuss three reference frames relevant to these processes. We
discuss underlying basic principles of such algorithms, their architecture,
and efficiency, and compare the outcomes of implementation of such algorithms
with the results of experiments performed on the human hand.
\end{abstract}

\tableofcontents

\section{Introduction: The Nervous System, Motor Control, and Stability Search
Problem}

Many functions of the central nervous system (CNS) can be described as
combining numerous elements (we will refer to their outputs as elemental
variables) into relatively low-dimensional sets related to such functions as
cognition, perception, and action. The existence of such low-dimensional sets
ensures stability of percepts, thoughts, and actions despite the variable
contributions from the elements (sensory receptors, neurons, motor units,
etc.) and changes in the environment. Here, we try to offer a mathematical
description of processes that could bring about such stability based on
variability. 
The aim of this paper is not to mimic a control algorithm for some particular
task, which the CNS solves, but rather to find a conceptual framework, though
vague at the moment, which would allow one to offer mathematical principles
that can shape neural activity associated with a variety of tasks.
We use, as an example,
the production of voluntary movements by redundant sets of elements. 
Our approach is not based on an a priori cost function, but on an intuitive, 
simple algorithmic principle. Of course, starting from a very complex algorithm 
may have an a priori advantage in achieving high probability of success as 
compared to simpler algorithms. We start building a model from a linear dynamic system and 
then add operational complexity as needed to achieve a reasonable success 
probability.
At any level of description, the system for movement production is apparently
redundant \cite{BernsteinNA1930}. This means that the number of its elemental state
variables is larger than the number of constraints associated with typical
tasks. Depending on the selected level of analysis, elements and elemental
variables are defined differently (reviewed in \cite{LatashML}). For
example, kinematic analysis of multi-joint action frequently considers
individual joint rotations as elemental variables. Kinetic analysis of the
force and moment production by parallel chains (such as the digits of the
human hand) may consider forces/moments produced by the digits as elemental
variables. Analysis of muscle activation may be based on firing patterns of
individual motor units as elemental variables. And so on. The apparent
redundancy typical of all these examples has been re-cast as abundance 
\cite{GelfandIM,LatashML(2012)} to emphasize that the extra elemental
variables are not the sources of computational problems for the CNS but an
essential component of the design that allows combining stability of actions
with flexibility of performance (changing actions, responding to
perturbations, performing several actions in parallel, etc.).

Consider the following example as an illustrations of our goal and
approach. Imagine a walking person who suddenly steps on a slippery
surface.The slip is typically followed by a very complex pattern of movements
of all body parts resulting in restoring balance in a large percentage of
cases. Each time a slip occurs the movement pattern looks unique. We assume
here that such highly variable patterns emerge as a result of a single,
relatively simple algorithm applied to cases with varying initial conditions
at the slip. What could be the search algorithm for new stability, presumably
formulated as a set of simple rules realized by the physical/physiological
system without explicit computational steps ? What is the success rate of not
falling yielded by such an algorithm ? We expect a balance between the
operational simplicity of the algorithm and the success probability. What
could be the relation of such an algorithm to established notions in the field
of motor control such as hierarchical control and uncontrolled manifold
hypothesis\cite{ScholzJP} ?

In the most simple mathematical setting, the equilibrium control problem 
(task)
 can be seen as a linear requirement $Y_{j}=\sum_{i=1}^{N}\alpha_{j,i}x_{i}$
imposed to $N$ elemental variables $x_{i}$ (with $i$ running from $1$ to $N$)
describing states of a redundant set of effectors, which depends on desired
position of the object given by a set of variables $Y_{j}$ (with $j$
running from $1$ to $M$). The task is to make the subspace determined by this
set of linear conditions mechanically stable, that is to find a proper
feedback matrix $\beta_{i,j}$, which via the equation
\begin{equation}
\overset{\centerdot}{x}_{i}=\sum_{j=1}^{M}\beta_{i,j}\left(  y_{j}
-Y_{j}\right)  \label{EQ1}
\end{equation}
relates the time derivatives $\overset{\centerdot}{x}_{i}$ of the elemental
variables and the deviations $\delta y_{j}=y_{j}-Y_{j}$ of instantaneous
coordinates $y_{i}$ of the controlled object from their desired positions
$Y_{j}$. One has to construct a simple algorithm of yielding such a feedback
matrix $\beta_{i,j}$ for an arbitrarily predetermined task matrix
$\alpha_{j,i}$.

If now one eliminates the physical aspect of the problem and treats the
variables $x_{i}$ and $y_{j}$ as certain brain variables responsible for the
control of a physical object, then the model begins to describe a purely
mental process that operates with some neural variables $x_{i}$ responsible
for the effector activity and the neural variables $y_{j}$ responsible for the
perception of the controlled object state, whereas the matrix $\alpha_{j,i}$
connects these two groups of variables.
Here, we suggest a sequence of general principles that may be seen
as extensions of the idea that biological systems are reasonably sloppy
\cite{LatashML(2008),LoebGE}. Under this expression, we mean that approximate solutions may
be generated for typical tasks as long as they lead to success with
acceptable probability. This notion is closely related to the notion
of satisficing coined by Herbert Simon \cite{citeREF}. 
This means that they do not solve problems exactly
but rather use simplified rules that produce solutions that are good enough
(e.g., successful most of the time). From the mathematical point of view, even
for a simple linear model, construction of a feedback algorithm is a nonlinear
process, which is meant to vary the matrix elements of $\beta_{i,j}$ depending
on the size and the evolution history of variables $x_{i}$. First, we suggest
a requirement, that the algorithm should rely, as much as possible, on local
actions, that is, on actions that change the way a neural variable $x_{i}$
participates in the feedback, such that the decision to undertake an action is
based on the actual and previous values $x_{i}(t<t_{actual})$ of this very
variable, whereas other variable do not affect the decision. However, such a
purely local algorithm may be incapable of providing good stability
properties, and thus one needs to involve non-local actions, when the
participation of one variable depends on the state and the evolution history
of other variables. In such a case, our second principle is that the number of
nonlocal interventions should be kept at a minimum. Third, we would like to
formulate the decision criteria in simple, intuitive terms. In this study, we
start with a rule "Act on the most nimble" (the AMN-rule), when changes in the
local parameters occur for the most unstable variable first.

Two more points should be discussed in the context of the requirements imposed
on the algorithm. (i)
It is desirable that the algorithm ensures robustness of the feedback and yields 
such $\beta_{i,j}$ that can be used across the task $\alpha_{j,i}$ within certain limits.
Formally this implies that the
spectrum $\left\{  \omega_{m}\right\}  $ of eigenvalues of the matrix
$\omega_{jj^{\prime}}=\sum_{i=1}^{N}\alpha_{j,i}\beta_{i,j^{\prime}}$ remains
stable (with negative real parts) even when the task matrix experiences a
perturbation $\delta\alpha_{j,i}$. (ii) It is hard to imagine a direct
experimental access to the assumed neural variables. 
Since the classical works by the group of Georgopoulos \cite{REF-1,ref-12}, studies 
of cortical neuronal population vectors have provided plenty of evidence 
that such vectors correlate with kinematic movement characteristics \cite{REF-2,ref-22}. 
We would like to emphasize, however, that such correlations are strongly 
dependent on the external conditions of the experiments and do not prove 
the physical meaning of the population vectors. In contrast, we are 
implying neural variables directly related to parameters related to 
setting a task in a multi-dimensional space of elemental variables and
participating in stabilization of its performance.
Fortunately, analysis of
the mathematical structure of the problem can rely on the noise statistics.
The reason for this is the fact that, for a rather general model of noise, one
can find a Gaussian distribution of the measurements in the $M$-dimensional
space of the observed variables $y_{j}$, while the eigenvectors of the
covariance matrix of this distribution coincides with the eigenvectors of
$\omega_{jj^{\prime}}$, and a certain relation among the eigenvalues can be deduced.

The "act on the most nimble" rule makes our model essentially non-linear. 
However, in-between successive actions on the most nimble variables, 
the dynamical system can be treated as linear and open, which implies 
its being subjected to noise, defined as uncontrollable small and random  
deviations of the coordinates from their averages over short time intervals. 
Though biological systems typically do not  manifest Gaussian noise statistics, 
the basic characteristics, such as noise covariance matrix, still can reveal 
important instantaneous characteristics of the matrix governing the linear 
part of dynamics, such as eigenvectors and, to a certain extent, eigenvalues, 
thus helping to compare calculated and observed results.  

\section{The Stability Search Algorithms, Dynamics Control, and the Effect of
Noise}

In this Section we discuss the main ideas underlying construction of the
stability search algorithms. We start by discussing the mathematical
complexity of the problem, and specific requirements for solving this problem
by an analog computer - a model for the CNS at this stage. In this context we
talk about local algorithms contrasting them to the non-local ones and
introduce a new basis of neural variables where such local algorithms can be implemented.

We continue by exploring the ingredients required for the stabilization
algorithm construction. By proposing a simple model of the control we specify
the problem in terms of a task and a feedback matrices, showing that a random
choice of the feedback has extremely low chances of yielding the required
stabilization. We next turn to the simplest local equilibrium search algorithm
- "act on the most nimble", which can considerably increase the chances of
stabilization gaining. This idea turns to be even more efficient when the
feedback matrix is not random, but is tailored in a way which we call
"generations" structure, -- where each one-dimensional subspace is coupled to
a multidimensional "generation" of variables, that all have typical velocities
("nimbleness") of the same order of magnitude. Each "generation" has
"nimbleness" considerably slower that for the previous one and much higher
than for the next "generation". This idea is further developed when we account
for the possibility of non-local actions, where each next generation, when
started at a relevant time scale to implement local operations, is also
allowed to inhibits activity of previous generations. This is followed by
analysis of the case where each generation nimbleness is a tunable fit
parameter, and a solution can always be found, for a system of any
dimensionality, although this solution rely on the typical velocities that may
differ by many orders of magnitude for the first and the last generations.

The next topic we discuss in this Section is the relations between noise and
stability, which later on will help us to gain a deeper insight into the
dynamics of the system by analyzing experimentally observed covariances of the
noise. The term \textquotedblleft noise\textquotedblright is ambiguous in
neurophysiology (reviewed in \cite{Stain}. For example, during quiet
standing, humans show spontaneous deviations of the body from the vertical
(postural sway), which are a superposition of two components, rambling and
trembling  \cite{ZatsiorskyVM(1999)}. The former reflects migration of the
equilibrium point with respect to which balance is organized, while the latter
reflects oscillations about that moving equilibrium point. Rambling has been
viewed as a reflection of an unintentional but meaningful search for limits of
stability and, hence, it would be counterintuitive to classify it as
\textquotedblleft noise\textquotedblright. Nevertheless, for the purposes of
the current study, we will use \textquotedblleft noise\textquotedblright to
imply spontaneous variance in physiological signals that is not related to the
explicit task and is not induced by an identifiable external perturbation. We
understand that this definition may attribute to noise physiologically
meaningful processes.

We conclude the Section by considering a realistic realization of the control
algorithm in hierarchical systems, show that for this architecture of the
feedback damping of the dynamic variables is indispensable and propose a model
of self-adjusting damping, which yields time dependencies of the dynamic
variables resembling ones observed experimentally.

\subsection{Control Complexity and Coordinate Frames}

Finding solution of $M$ linear equations $Y_{j}=\sum_{i=1}^{N}\alpha
_{j,i}x_{i}$ for $N>M$ variables is a problem of a polynomial complexity $\sim
M^{2}$. It is sufficient to take the first $M$ columns of the $M\times N$
matrix $\alpha_{i,j}$ and invert the resulting $M\times M$ square matrix,
provided it is not degenerate. Solutions of the problem for the first $M$
variables $\left\{  x_{1}\ldots x_{M}\right\}  $ form a $M$-dimensional
subspace $S_{M}$ in the entire $N$-dimensional space, parametrized by the
remaining $N-M$ variables $\left\{  x_{M+1}\ldots x_{N}\right\}  $.
Construction of the inverse matrix $\alpha_{i,j}^{-1}$ based on the sequential
finding of its line vectors complemented by the orthonormalization of these
vectors with those found earlier indeed implies the number of operations on
the order of $M^{2}$.

To solve this problem on an analog computer, one should construct and
implement a dynamic process that has the $M$-dimensional subspace $S_{M}$ as a
stable stationary manifold, as suggested by Eq.(\ref{EQ1}). Apparently, the
additional requirement of stability augments the mathematical complexity of
the relevant algorithm, but once this requirement is fulfilled, the result
offers an advantageous generalization -- it can be implemented for solving
systems of nonlinear equations that is a problem of a much higher complexity.
Here, we are going to propose several strategies of constructing heuristic
algorithms that result in the dynamic stabilization of subspaces defined by
systems of $M$ equations for $N>M$ variables.

Two points have to be elucidated. The first is that, according to the general
vision, an analog computer is a collection of rather simple, although linearly
interacting, dynamic systems - elements, while the algorithms are viewed in
this context as sequences of instrumental prescriptions of how to modify these
interactions. Each of the prescribed conditions is imposed on the value of a
dynamic variable describing the state of one element and the prescribed action
is applied to the same or to another element. Prescribed action on one chosen
element that depends only on the state of either that element or on another
chosen element implies so-called locality or bi-locality, respectively, of the
elementary step of the algorithms. The second point is that nonlinearity
of the analog computer generated by such a local and/or bi-local algorithm may
not result in universal stability of the dynamics. In other words, for some
initial conditions the system may reach a stationary solution, while for other
initial conditions it becomes unstable. In such a situation, one can speak
about the success rate $R(N,M)$ of the algorithm, and this statistical
quantity serves as its quality characteristic.

We now move to implementing the two principles mentioned in the Introduction:
(1) Reliance primarily on local actions while nonlocal actions are minimally
involved; and (2) Using the intuitive AMN-rule. First, we introduce three
types of specific bases related to certain linear combinations of the body
variables. First, there is a \emph{measurement} basis formed by experimentally
accessible variables, for example positions, forces, or muscle activations.
Second, there is another, task-dependent, basis formed by
so-called\emph{ modes}, that are linear combinations of the former variables
independently fluctuating under the action of noise and coinciding with the
eigenvectors of the aforesaid matrix $\omega_{jj^{\prime}}$ 
(cf. \cite{DanionF,KrishnamoorthyV}. The modes are expected to be stable for
broad ranges of tasks and not to change without specialized training. The
third basis is the reference system relying on the variables $x_{i}$
responsible for the control over the body state. To our knowledge, this
coordinate system has not been defined previously. Each of the coordinates of
the third system represents combinations of modes that are task specific and
relatively quickly adjustable to changes in the external conditions of task
execution, for example to changes in stability requirements (e.g. \cite{Asaka}).
This basis comprises the variables that experience local control,
that is, changes in the feedback loop gain with respect to each of these
variables $x_{i}$ produced once a certain functional $z_{i}\left\{
x_{i}(t)\right\}  $of these very variable reaches a critical value. Such gain
changes may occur gradually with changes in the value of the functional. To
summarize, the task is considered within a preexisting, task-dependent basis.

\subsection{The Control Structure}

To solve the problem of a subspace stabilization one first takes an
arbitrarily $M\times N$ matrix $\widehat{\alpha}$ that imposes conditions of
Eq.(\ref{EQ1}) with $Y_{i}=0$ thus defining the required subspace
$S_{M}=\left\{  y_{i}=0\right\}  $. Thus the task is to find a $N\times M$
linear feedback matrix $\widehat{\beta}$ that, by determining the derivatives
\begin{equation}
\frac{d}{dt}x_{i}=\sum_{j=1}^{M}\beta_{i,j}y_{j}, \label{EQ3}%
\end{equation}
prescribes the modifications of $\left\{  x_{i}\right\}  $ leading towards the
subspace $S_{M}$. Dynamics of the system is therefore given by the equation
\begin{equation}
\frac{d}{dt}x_{i}=\sum_{j=1}^{N}\sum_{k=1}^{M}\beta_{i,k}\alpha_{k,j}x_{j}.
\label{EQ4}%
\end{equation}
and a stable $M$-dimensional subspace $S_{M}$ exists in the $N$-dimensional
space if all the non-zero eigenvalues of the degenerate square $N\times N$
matrix $\widehat{\Omega}=\widehat{\beta}\widehat{\alpha}$ of the rank $M$ have
negative real parts. The probability to satisfy this requirement for randomly
chosen either $\widehat{\alpha}$ or both $\widehat{\alpha}$ and $\widehat
{\beta}$ is rather low, and in order to stabilize $S_{M}$ with a high enough
success rate $R(N,M)$ for a chosen $\widehat{\alpha}$, a randomly or
systematically chosen linear feedback matrix $\widehat{\beta}%
_{\mathrm{initial}}$ may need to be modified $\widehat{\beta}%
_{\mathrm{initial}}\rightarrow\widehat{\beta}_{\mathrm{modified}}$ once it
does not ensure negativity of the $\widehat{\beta}\widehat{\alpha}$
eigenvalues real parts and hence does not yield the required stabilization. A
sequence of the modifications $\widehat{\beta}_{\mathrm{modified(n)}%
}\rightarrow\widehat{\beta}_{\mathrm{modified(n+1)}}$ can be continued
according to a certain algorithm until stabilization is achieved.

Each elementary act of the modification algorithm rely on three main
ingredients: (i) the variable $x_{i}$ initiating the action, (ii) type of the
action itself $\widehat{\beta}_{\mathrm{modified(n)}}=\widehat{\mathbf{G}%
}\widehat{\beta}_{\mathrm{modified(n+1)}}$ implementing the modification
$\widehat{\beta}_{\mathrm{modified(n)}}\rightarrow\widehat{\beta
}_{\mathrm{modified(n+1)}}$,which is specified by a non-linear functional
matrix $\widehat{\mathbf{G}}(\left\{  x_{f}(t)\right\}  ,\left\{
x_{i}(t)\right\}  )$, and (iii) the variable $x_{f}$ subject to the action.
Once the source and the target variables coincide, we encounter a so-called
local algorithm, which implies that for $f\neq i$ we deal with a non-local
elementary act of the algorithm. During last decades, the difference between
local and non-local actions are has been widely discussed in the context of
Quantum Informatics, where physical implementation of non-local
transformations, so-called quantum gates, is a much more challenging
experimental task as compared to local transformations. In a situation where
each variable $x_{k}$ is associated with some location in space, the situation
typical of Quantum Informatics becomes generic: Local algorithms are much
easier to implement in a physical realization of the controlled systems than
non-local ones since they do not require transportation of the nonlinear
action from one spatial location to another.

\subsubsection{No algorithm}

We first consider the simplest realization of an analog computer working
without a controlling algorithm. For the sake of presentation simplicity,
assume that the desired subspace $S_{M}$ corresponds  to zero values of all
the variables $y_{i\in\left\{  1,M\right\}}$. Non-zero values of these
variables will thus be employed as entries into the feedback loop given by a
rectangular $N\times M$ matrix $\beta_{i,j}$, such that Eq.(\ref{EQ3}) holds,
while Eq.(\ref{EQ4}) describes the system dynamics.

For a generic rectangular randomly chosen matrix $\alpha_{i,j}$ and the
randomly chosen feedback matrix $\beta_{i,k}$ the success rate $R(N,M)$ found
numerically scales approximately as $2^{-M}$, which is consistent with the
assumption that all non-zero eigenvalues of the matrix $\widehat{\Omega}$ may
have the real parts both negative and positive with equal probability and,
roughly speaking, different eigenvalues are statistically independent one from
one another.

\subsubsection{Local algorithm for random feedback}

Our aim now is to construct a control algorithm, which is capable of improving
the success rate. We call this algorithm the AMN-rule later. We begin with the
simplest case of a local algorithm by assuming that the feedback sign for a
given $x_{i}$ changes once a positive quantity
\begin{equation}
z_{i}(t)=\int_{0}^{t}\left(  \frac{dx_{i}}{dt}\right)  ^{2}dt \label{EQ5}%
\end{equation}
exceeds a threshold value $Z$. This represent an example of "act on the most
nimble" control, where, as a first guess, we consider the integral of the
rates squared, which is an analog of the strictly positive dissipated heat
functional typical of electrical circuits. Further, we will consider other
strictly positive functionals. Formally, the local control implies that the
set of equations (\ref{EQ4}) is modified by the presence of a sign function
and reads
\begin{align}
\frac{d}{dt}x_{i}  &  =G_{ii}\sum_{j=1}^{N}\sum_{k=1}^{M}\beta_{i,k}%
\alpha_{k,j}x_{j}\label{EQ6.1}\\
G_{ii}  &  =\mathrm{sign}(Z-z_{i}(t)). \label{EQ6.2}
\end{align}

This strategy yields $100\%$ success rate for the case of $M=1$, that is the
case where the matrix $\widehat{\beta}\widehat{\alpha}$ has rank 1. In this
case the dynamic process occurs in a subspace of the dimension $1$ such that
the dynamic equation
\begin{equation}
\frac{d}{dt}y_{1}=\sum_{i=1}^{N}\alpha_{1,i}G_{ii}\beta_{i,1}y_{1} \label{EQ6}
\end{equation}
for the variable $y_{1}=\sum_{i=1}^{N}\alpha_{1,i}x_{i}$ determines time
dependencies of all the other variables $x_{i}=\beta_{i,1}y_{1}$. Indeed, if
the quantity $\omega=\sum_{i=1}^{N}\alpha_{1,i}\beta_{i,1}$ is positive,
$y_{1}$ exponentially increases in time. This means that the variable $x_{i}$
corresponding to the largest $\left\vert \beta_{i,1}\right\vert $ produces the
highest "Joule's heat" $z_{i}(t)$ and at some moment of time, when this
quantity exceeds the threshold value $Z$, the sign of $\mathrm{sign}
(Z-z_{i}(t))\beta_{i,1}$ changes. For a positive $\alpha_{1,i}$, this results
in a decrease of $\omega$ and slowing-down of the instability. For a random
matrix $\widehat{\alpha}$, however, the matrix element $\alpha_{1,i}$ can
equally be negative or positive; in the latter case, the change of the sign
results in the instability speeding-up. In this case, the exponential growth
of the variables $x_{i}$ continues, and after awhile, the next biggest
$\left\vert \beta_{i,1}\right\vert $ leads to a change of the corresponding
feedback sign. Changes of the signs continue till $\omega$ becomes negative,
and the dynamic process becomes stable. This will definitely occur for the
matrix $\widehat{\beta}$ of the rank $1$, but for higher ranks
this stabilization may not happen. Moreover, numerical search shows that the
success rate $R_{\mathrm{local}}$ of such stabilization drops with the
increasing rank $M$. The results of the numerical simulation for this case are
shown in Table \ref{table1}, the fourth row ($R_{\mathrm{local}}$).
\begin{table*}[ht]
\[
\begin{array}[c]{|c|ccccccc|}
\hline
M & 1 & 2 & 3 & 4 & 4 & 4 & 5\\
N & 10 & 20 & 30 & 15 & 40 & 60 & 30\\
\hline
R_{\mathrm{random}} & 0.50 & 0.24 & 0.14 & - & 0.06 & - & -\\
R_{\mathrm{local}} & 1 & 0.77 & 0.46 & 0.17 & 0.16 & 0.154 & 0.06\\
R_{\mathrm{tailored}} & 1 & 0.85 & 0.62 & 0.32/0.28 & 0.36 & 0.31 &
0.25/0.21\\
R_{\mathrm{generations}} & 1 & 0.9 & 0.68 & 0.42/0.46 & 0.66 & - & 0.45/0.31 \\
\hline
\end{array}
\]
\caption{The success rate of various algorithms stabilizing $M$-dimensional subspace of $N$-dimensional space.}
\label{table1}
\end{table*}
These results have to be compared with $R_{\mathrm{random}}$ for the not
controlled random feedback. While the success rate is higher for the
controlled system for all $M$ and $N$, it drops rather quickly with the rank
$M$, while being relatively less sensitive to $N$ (compare the $R$ values for
$M=4$ and $N=15$, $40$, and $60$).

\subsubsection{Tailored linear feedback and the local algorithm}

Remaining within the general case of Eq. (\ref{EQ6.1}), we replace the random
feedback matrix $\beta_{i,j}$ by another one, constructed to augment the
success rate for higher $M$.We take it in the form%
\begin{equation}
\widehat{\beta}=\left(
\begin{array}
[c]{ccccc}%
\overrightarrow{v} & 0 & 0 & \ldots & 0\\
0 & q\overrightarrow{v} & 0 & \ldots & 0\\
0 & 0 & q^{2}\overrightarrow{v} & \ldots & 0\\
\ldots & \ldots & \ldots & \ldots & \ldots\\
0 & 0 & 0 & \ldots & q^{M-1}\overrightarrow{v}%
\end{array}
\right)  \label{EQ7}%
\end{equation}
where $\overrightarrow{v}$ is a column vector of the size of the ratio $N/M$,
which we assume to be an integer. Here $q$ is a small parameter, which ensures
that at each time scale $\sim q^{-n}$, one has to deal with a subspace of rank
just $1$ with the same algorithm as for the case of $M=1$, while the
subspaces corresponding to the larger matrix elements $\sim q^{-k<n}$ are
assumed to have already been stabilized earlier.

In simple words the main idea is: there are generations of elemental variables
organized by their expected nimbleness, which scales as $q^{g}$, that is,
exponentially with the number $g$ of the generation. The controller uses the
AMN-rule within an appropriate channel (which is a set of variables $x_{i}$
grouped by their common input from a component of the vector $\overrightarrow
{y}$) and then deals with the next most nimble channel without returning to
the first one, and so on. This approach improves the success rate for higher
$M$; however, even it cannot guarantee $100\%$ convergence. The results of the
numerical work for the success rate $R_{\mathrm{tailored}}$ in the case $M=3$,
$N=9$, $q=0.07$,
\begin{equation}
\overrightarrow{v}=\left(
\begin{array}[c]{c}
1\\
1.5\\
2
\end{array}
\right)
\label{EQ8}
\end{equation}
are shown on the fifth line of Table \ref{table1}, while the results for $q=0.1$
calculated for two of these cases are shown after the slash.

\subsubsection{Tailored linear feedback and the generation-specific non-local
algorithm}

The idea of bi-local control allows excluding undesired changes in the
feedback sign determined at an earlier stage of control that may be induced at
a later stage. In the feedback matrix Eq.(\ref{EQ7}), we identify parts that
belong to different generations, corresponding to different orders of the
parameter $q$. The first generation corresponds to $q^{0}$, and the last, the
most recent, generation to $q^{M-1}$. Evidently, each generation accounts
for the feedback at the corresponding time scale. The idea of the control
algorithm is that changing the sign of a variable belonging to a generation
blocks changes of the signs of the variables belonging to all former generations.

Formally, the bi-local control implies that the set of equations
(\ref{EQ5},\ref{EQ6.1}) is modified by the presence of step factors
$\Theta(Z-z_{i})$ at the derivatives for the variables $z_{i}$ corresponding
to "generations" that happened after than that of $i$, and reads%
\begin{align}
\frac{d}{dt}z_{i}(t)  &  =\left[  \prod\limits_{r\in\mathrm{inferior}%
(i)}\Theta(Z-z_{r}(t))\right]  \left(  \frac{dx_{i}}{dt}\right)
^{2}\label{EQ9}\\
\frac{d}{dt}x_{i}  &  =G_{ii}\sum_{j=1}^{N}\sum_{k=1}^{M}\beta_{i,k}%
\alpha_{k,j}x_{j}\nonumber\\
G_{ii}  &  =\mathrm{sign}(Z-z_{i}(t)),\nonumber
\end{align}
where $\Theta(x)$ is the step function. The first equation of this system
shows that the functionals $z_{i}(t)$ governing the feedback signs are no
longer local, since dynamic equations ruling these quantities depend not only
on the corresponding local squared velocities but also on the values of the
functionals for other variables. This construction further improves the
success rate. The results of the numerical work for the success rate
$R_{\mathrm{generations}}$ in this case are shown in the last line of the
Table \ref{table1} for $q=0.07$-$0.1$.
\begin{figure}[ht]
\begin{center}
\includegraphics[width=3in]{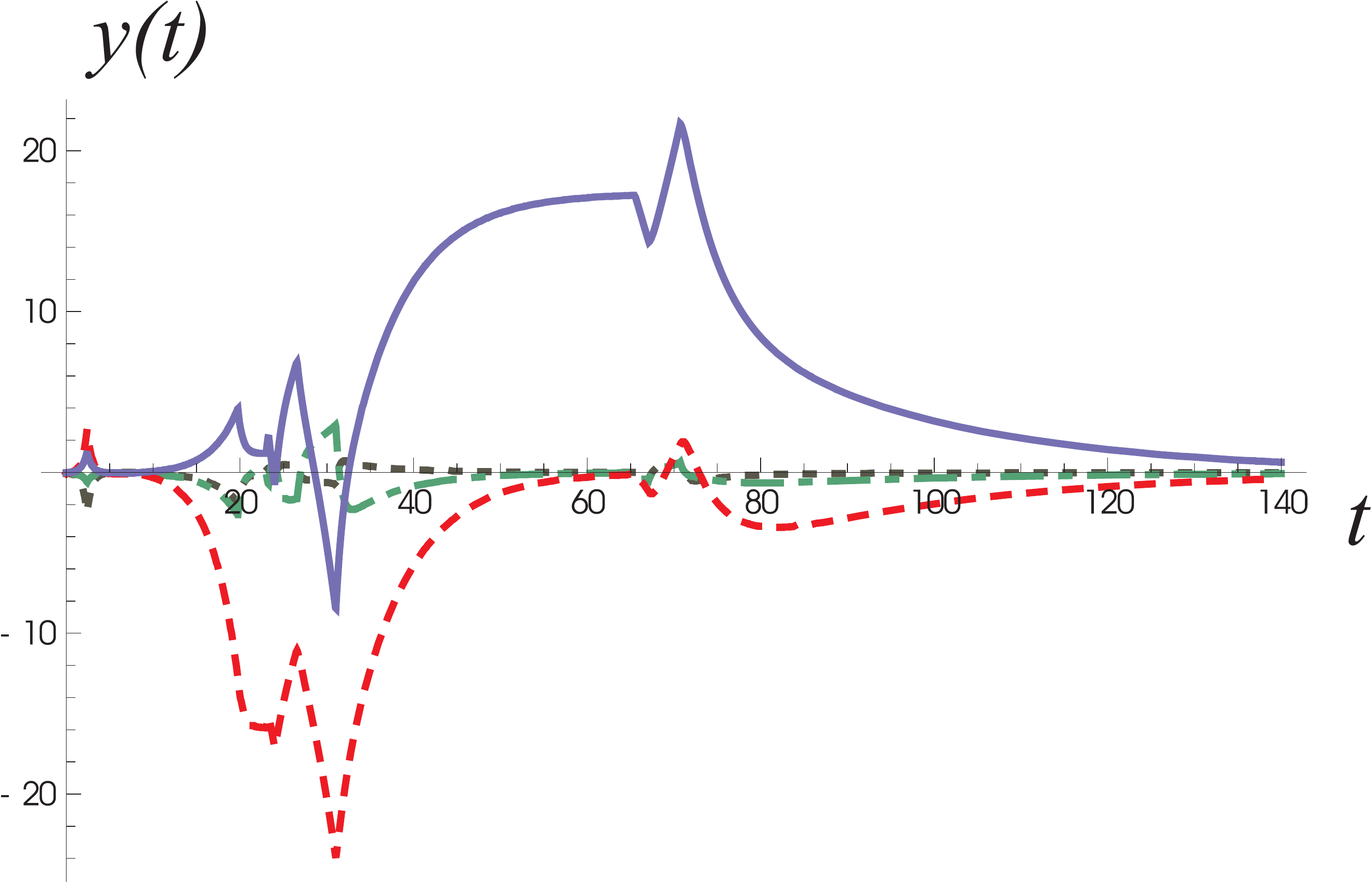}
\caption{An example of stabilizing a four-dimensional subspace in a
$40$-dimensional space. Time dependencies of the parameters $y_{1\ldots4}$,
that determine the subspace for $y_{i}=0$ are shown. The first variable
$y_{1}$ (the blue solid line) corresponds to the first generation of the
variables $x_{1}\ldots x_{10}$ having the strongest coupling. The second
variable, $y_{2}$, (the red dashed line) is coupled to the variables
$x_{11}\ldots x_{20}$ with weaker coupling constants, scaled by the factor
$q=0.4$ with respect to the first variable. The third and the fourth
generations (the dash-dot and the dotted lines, green and black, respectively)
have couplings scaled by the factors $q^{2}$ and $q^{3}$, respectively. In the
figure, one can see seven sequential discontinuities of the derivatives of
the dependencies related to the changes of signs of the most nimble variables
$x_{i}$, that first occurs in the first generation, then in the second, etc.
The changes in each of the generations manifest themselves in all dependencies
via the matrix $\widehat{\alpha}$}.
\label{FIG1}
\end{center}
\end{figure}

In Fig.\ref{FIG1} we show an example of the time dependent deviations
$y_{1\ldots4}(t)$ in the case of a successful control for $N=4$, $M=40$,
$q=0.4$, where the abrupt changes of the dependencies reflect changes of the
feedback signs.

\subsection{Feedback Strength Exploring Algorithm}

The bi-local algorithm can be modified to gain the $100\%$ success rate if
instead of the fixed power factors $q^{n}$ entering Eq.(\ref{EQ7}) one allows
sequential choosing of these feedback strength parameters for each next
generation. In a sense, this algorithm implies learning, that is, given a task
$\widehat{\alpha}$ , it modifies the feedback matrix $\widehat{\beta}$ once
the sign-changing algorithms does not result in the required subspace
stabilization. More specifically, given $q<1$, similar to Eq.(\ref{EQ7}) where%
\begin{equation}
\widehat{\beta}=\left(
\begin{array}
[c]{ccccc}%
\overrightarrow{v} & 0 & 0 & \ldots & 0\\
0 & q^{p_{1}}\overrightarrow{v} & 0 & \ldots & 0\\
0 & 0 & q^{p_{2}}\overrightarrow{v} & \ldots & 0\\
\ldots & \ldots & \ldots & \ldots & \ldots\\
0 & 0 & 0 & \ldots & q^{p_{M-1}}\overrightarrow{v}%
\end{array}
\right)  , \label{EQ.9.0}%
\end{equation}
the power parameters $p_{1}\leq p_{2}\leq\ldots p_{M-1}$  that have to be
chosen such that the subspace $\left\{  y_{i}=0\right\}  $ is stable.

One begins with all $p_{i}=\infty$, that is, with a one-dimensional subspace,
which can always be made stable by proper choice of signs. Next, one sets
$p_{1}$ to zero, and starts to implement the sign changing algorithm in the
subspace of the second diagonal cell of Eq.(\ref{EQ10}). If this algorithm
does not lead to a stable subspace of the dimension $2$, one increases $p_{1}$
by unity, and implements the sign-changing algorithm for the second cell once
again. Repeating this procedure leads to finding $p_{1}$ such that the
two-dimensional space becomes stable. Next, one turns to the third cell of the
matrix Eq.(\ref{EQ10}), puts $p_{2}=p_{1}$, and implements the sign-changing
algorithm complemented by augmentation of $p_{2}$ by unity, if needed, until
the third dimension gets stable. The procedure being sequentially applied to
all the cells of Eq.(\ref{EQ10}). finally yields a feedback matrix
$\widehat{\beta}$ stabilizing the required subspace of the dimension $M$. for
the chosen task matrix $\widehat{\alpha}$.

\begin{figure}[ht]
\begin{center}
\includegraphics[width=3in]{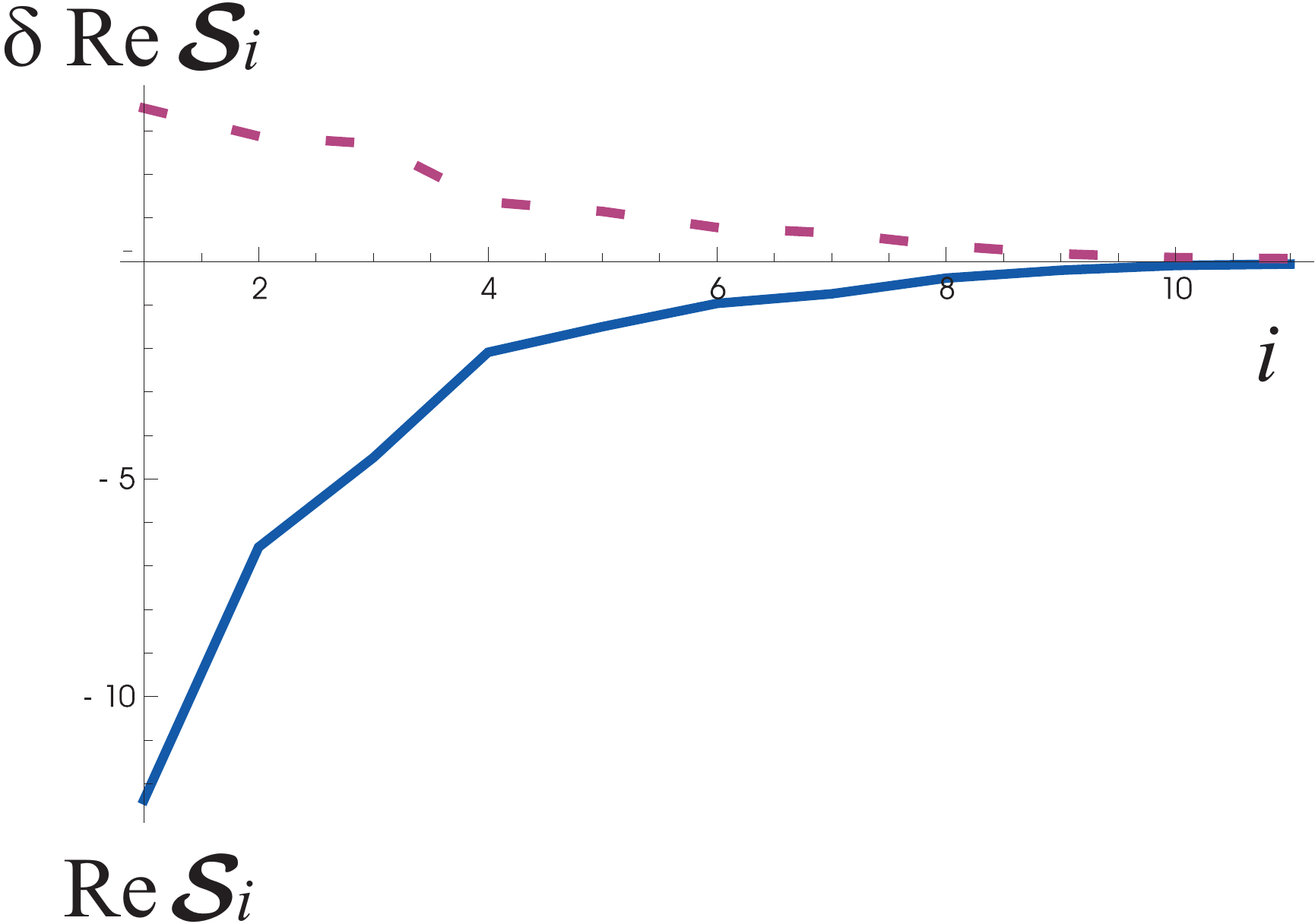}
\caption{The average sorted eigenvalues (solid line) and their average mean
absolute value deviation (dotted line) obtained for the case $N=143,M=11$ with
the help of the feedback strength-exploring algorithm. The average has been
taken over $25$ casts of the random matrix $\widehat{\alpha}$.}
\label{FIG2}
\end{center}
\end{figure}

For the case of $N=100$, $M=10$, in Table \ref{table2}
\begin{table*}[ht]
\[
\begin{array}[c]{|c|cccccccccc|}
\hline
\left\{  \mathcal{S}_{i}\right\}  _{m}\setminus i & 1 & 2 & 3 & 4 & 5 & 6 &
7 & 8 & 9 & 10\\
\hline
\left\{  \mathcal{S}_{i}\right\}  _{1} & 1.59 & 1.57 & 1.57 & 0.00253 &
0.00253 & 0.00161 & 0.00161 & 0.000144 & 0.000144 & 2.0\;10^{-5}\\
\left\{  \mathcal{S}_{i}\right\}  _{2} & 1.2 & 0.219 & 0.219 & 0.197 & 0.197 &
0.187 & 0.0214 & 0.0214 & 1.69\;10^{-3} & 1.69\;10^{-3}\\
\left\{  \mathcal{S}_{i}\right\}  _{3} & 0.849 & 0.131 & 0.131 & 0.0297 &
0.0297 & 0.0223 & 0.0223 & 1.51\;10^{-4} & 2.3\;10^{-5} & 2.3\;10^{-5}\\
\left\{  \mathcal{S}_{i}\right\}  _{4} & 2.86 & 1.28 & 1.28 & 0.317 & 0.317 &
0.0807 & 0.0807 & 5.75\;10^{-3} & 5.75\;10^{-3} & 2.26\;10^{-3}\\
\left\{  \mathcal{S}_{i}\right\}  _{5} & 2.27 & 0.37 & 0.37 & 0.0824 &
0.0824 & 2.46\;10^{-3} & 8.38\;10^{-4} & 8.38\;10^{-4} & 1.24\;10^{-4} &
6.33\;10^{-5}\\
\left\{  \mathcal{S}_{i}\right\}  _{6} & 1.48 & 0.614 & 0.614 & 0.411 & 0.13 &
0.13 & 0.109 & 0.109 & 0.1 & 0.1\\
\left\{  \mathcal{S}_{i}\right\}  _{7} & 3.54 & 3.54 & 1.21 & 1.21 & 0.208 &
0.208 & 0.0317 & 0.0317 & 2.87\;10^{-3} & 2.87\;10^{-3}\\
\left\{  \mathcal{S}_{i}\right\}  _{8} & 6.58 & 0.751 & 0.27 & 0.27 & 0.187 &
0.187 & 0.0363 & 0.0363 & 0.012 & 0.012\\
\left\{  \mathcal{S}_{i}\right\}  _{9} & 1.85 & 1.85 & 0.507 & 0.507 &
0.0478 & 0.0478 & 0.0244 & 0.0244 & 6.83\;10^{-4} & 6.83\;10^{-4}\\
\left\{  \mathcal{S}_{i}\right\}  _{10} & 0.652 & 0.652 & 0.602 & 0.602 &
0.138 & 0.138 & 0.0734 & 0.0734 & 4.71\;10^{-3} & 4.71\;10^{-3} \\
\hline
\end{array}
,
\]
\caption{Sets of the negative real parts of the eigenvalues of matrix $\hat\alpha \hat\beta$  
obtained with the help of the feedback strength exploring algorithm for ten randomly 
chosen task matrices $\hat\alpha$.
}
\label{table2}
\end{table*}
we present sets $\left\{  \mathcal{S}_{i}\right\}  _{m}$ of the negative real
parts $\mathcal{S}_{i}$ of the eigenvalues $c_{i}$ of the matrix
$\widehat{\alpha}\widehat{\beta}$ obtained after $10$ castings ($m=1\ldots10$)
for random matrices $\widehat{\alpha}$ and matrices $\widehat{\beta}$ found as
a result of implementation of the described algorithm. In Fig.\ref{FIG2} we
show the average eigenvalues and their mean absolute value deviation from the
average calculated with the help of this algorithm. Both these quantities are
exponentially decreasing with the eigenvalue number. Note that it does not
make sense to put the results of the calculations into Table \ref{table1}, since the
corresponding entries are all unities.

This result means that the local control based on the change of the feedback
sign for the most nimble variable (the AMN-rule) combined with a simple
non-local control solves the feedback search problem. This nonlocal control
requires freezing the feedback parameters for the previous generations once a
new generation starts to perform the AMN-rule-based local control. Moreover,
it requires a learning procedure with adjusted attenuation of the generation
couplings. Though the proposed algorithm solves the problem for any large
dimension $M$, and thus explicitly demonstrates a possibility to construct a
$100\%$ successful algorithm, it may yield the coupling strengths so weak,
that the obtained solution has no practical value.

\subsection{Susceptibility to Noise Reveals Feedback Structure \label{NOISE}}

We now address the question: What happens with the convergence towards the
stabilized subspace ensured by the feedback matrix
\begin{equation}
\Omega_{i,j}=\mathrm{sign}_{i}(t\rightarrow\infty)\sum_{k=1}^{M}\beta
_{i,k}\alpha_{k,j} \label{EQ9.1}%
\end{equation}
in the presence of noise ? Again, as earlier, we emphasize that for
physiological systems "noise" implies external task-independent actions that
can be of various origin, including both spontaneous changes in intrinsic
neural variables and uncontrolled action of an external force field. Here
$\mathrm{sign}_{i}(t\rightarrow\infty)$ denotes the sign, which results from
successful implementation of the control algorithm. We specify this question
by asking: How far from the average positions $x_{i}(t)$ satisfying the
dynamic equations can the actual variables $X_{i}(t)=x_{i}(t)+\delta x_{i}(t)$
deviate in the presence of a time-dependent noise $f_{i}(t)$ ? It can be
immediately answered in the basis of the eigenvectors $\overrightarrow
{\mathcal{X}}_{i}$ of $\Omega_{i,j}$ where $\overrightarrow{x}(t)=\sum
_{i}x_{i}(t)\overrightarrow{\mathcal{X}}_{i}$.

One considers the dynamic equations
\begin{equation}
\frac{d}{dt}\delta x_{i}(t)=\mathcal{S}_{i}\delta x_{i}(t)+f_{i}(t),
\label{EQ10}%
\end{equation}
for the deviations $\delta x_{i}(t)$, where $\mathcal{S}_{i}$ denotes
eigenvalues of $\widehat{\Omega}$. The solution
\begin{equation}
\delta x_{i}(t)=\int\limits_{0}^{t}e^{\mathcal{S}_{i}(t-\tau)}f_{i}(\tau
)d\tau, \label{EQ11}%
\end{equation}
of this equations in the Fourier representation
\begin{equation}
\delta x_{i}(\omega)=\frac{1}{\mathcal{S}_{i}-i\omega}f_{i}(\omega),
\label{EQ12}%
\end{equation}
suggests that the spectral noise intensity $\left\vert x_{i}(\omega
)\right\vert ^{2}$ of the coefficients $x_{i}(t)$ is related to the spectral
noise intensity $\left\vert f_{i}(\omega)\right\vert ^{2}$ of $f_{i}(t)$ via
the relation
\begin{equation}
\left\vert x_{i}(\omega)\right\vert ^{2}=\frac{1}{\left\vert \mathcal{S}%
_{i}-i\omega\right\vert ^{2}}\left\vert f_{i}(\omega)\right\vert ^{2},
\label{EQ13}%
\end{equation}
which corresponds to the susceptibility $\left\vert \mathcal{S}_{i}%
-i\omega\right\vert ^{-2}$.

In order to find the net mean square deviation $\sqrt{\left\langle \delta
x_{i}^{2}\right\rangle}=\sqrt{\int d\omega\left\vert \delta x_{i}%
(\omega)\right\vert ^{2}/2\pi}$ of the variable $X_{i}$ , one has to choose a
model for the random perturbing forces or noise. In contrast to the Gaussian
noise, typical of physical multi-body systems, due to the central limit theorem
of statistics, biological systems may, and usually do manifest other type of
noisy behavior, since the underlying processes may emerge from complicated
nonlinear and non-random processes developing at the time scales much shorter
as compared to the time scale of the process under consideration. The simplest
model includes randomly chosen static forces $f_{i}$ acting on the relevant
variables during a time interval $T$ and, in each, next time intervals, these
forces show random values. This yields
\begin{equation}
\left\vert f_{i}(\omega)\right\vert ^{2}=\left(  \frac{\sin\frac{\omega T}{2}%
}{\omega T}\right)  ^{2}\left\vert f_{i}\right\vert ^{2}, \label{EQ14}%
\end{equation}
and hence
\begin{eqnarray}
\left\langle \delta x_{i}^{2}\right\rangle &= &\left\vert f_{i}\right\vert
^{2}\int\frac{d\omega}{2\pi}\frac{1}{\left\vert \mathcal{S}_{i}-i\omega
\right\vert ^{2}}\left(  \frac{\sin\frac{\omega T}{2}}{T\omega}\right)
^{2} \nonumber \\
&= &\left\vert f_{i}\right\vert ^{2}\frac{\operatorname{Re}\left[  \left(
e^{\mathcal{S}_{i}T}-1\right)  \mathcal{S}_{i}^{-2}\right]}
{T\operatorname{Re}\mathcal{S}_{i}}. \label{EQ15}
\end{eqnarray}
Average square variances $\left\langle \delta x_{i}^{2}\right\rangle $ of the
independent modes correspond to eigenvalues $\mathcal{C}_{i}$ of the
covarience matrix, when one considers the problem in a basis, different from
$\left\{  \overrightarrow{\mathcal{X}}_{i}\right\}  $.

For the case $\mathcal{S}_{i}=0$ one should take into account a finiteness of
the evolution time interval $\mathcal{T}$ and replace $\mathcal{S}_{i}$ by
$i/\mathcal{T}$, thus obtaining the net mean squared deviation diffusively
rising with time%
\begin{equation}
\left\langle \delta x_{i}^{2}\right\rangle \sim\mathcal{T}/T.
\label{EQ16}
\end{equation}

Noise analysis may serve as a powerful tool of revealing the eigenvectors of
the matrix $\Omega_{i,j}$ of Eq.(\ref{EQ9.1}), which via the
Eqs.(\ref{EQ13},\ref{EQ14}), also gives some ideas about magnitudes of the
corresponding eigenvalues for a reasonably chosen model of the noise. One
needs to calculate the noise covariance matrix $C_{ij}=\left\langle \delta
x_{i}\delta x_{j}\right\rangle $ from the experimentally observed deviations
of the running measured values from their average values in the stationary
regime. Eigenvectors $\overrightarrow{\mathcal{X}}_{i}$ of this matrix should
coincide with those of $\widehat{\Omega}=\widehat{G}(t\rightarrow
\infty)\widehat{\beta}\widehat{\alpha}$. Moreover, this method can also be
employed for determining the eigenvalues of the time-dependent values of the
feedback matrix $\widehat{\Omega}(t)=\widehat{G}(t)\widehat{\beta}%
\widehat{\alpha}$, provided the time intervals where this object changes
significantly are long enough and exceed a typical correlation time of the noise.

Note that for the case $N>M$, the non-zero eigenvalues of the $N\times N$
matrix $\widehat{\Omega}=\widehat{G}\widehat{\beta}\widehat{\alpha}$ coincide
with $M$ eigenvalues of the $M\times M$ matrix $\widehat{\omega}%
=\widehat{\alpha}\widehat{G}\widehat{\beta}$. The remaining $N-M$ eigenvalues
are zeros, unless an additional requirement is imposed on the matrix
$\widehat{\Omega}$. Spread of the variations $\delta x_{i}^{2}$ in the
directions of the corresponding eigenvectors are expected to be large and be
governed by the spectral properties of the noise. In other words, given a task
matrix and a feedback matrix, which has been found with the help of the
stabilization search algorithm, there exist two subspaces in the space of the
variables $x_{i}$, the so-called uncontrolled manifold (UCM)\cite{ScholzJP} 
and its orthogonal complement (ORT) also sometimes addressed
as the null-space and the range-space, respectively. ORT is the task-specific
subspace expected to show high stability, which implies that variance in ORT
is expected to be small. In the UCM, variance is generally expected to be
large and increasing with time, unless there are other factors, outside the
explicit task formulation, that keep it within a certain range. In fact, after
a person becomes an expert in a motor task, repetitive attempts at the task
lead to variable combinations of elemental variables that all lead to
successful task performance within a permissible error margin (\cite{BernsteinNA1930};
reviewed in \cite{LatashML(2008)}). However, not all possible combinations of the
involved elemental variables are used across repetitive attempts. Such
self-imposed additional constraints may be addressed as 'perfectionism'; they
may reflect optimization with respect to a cost function (e.g., \cite{TerekhovAV}).
\begin{figure}[ht]
\begin{center}
\includegraphics[width=3in]{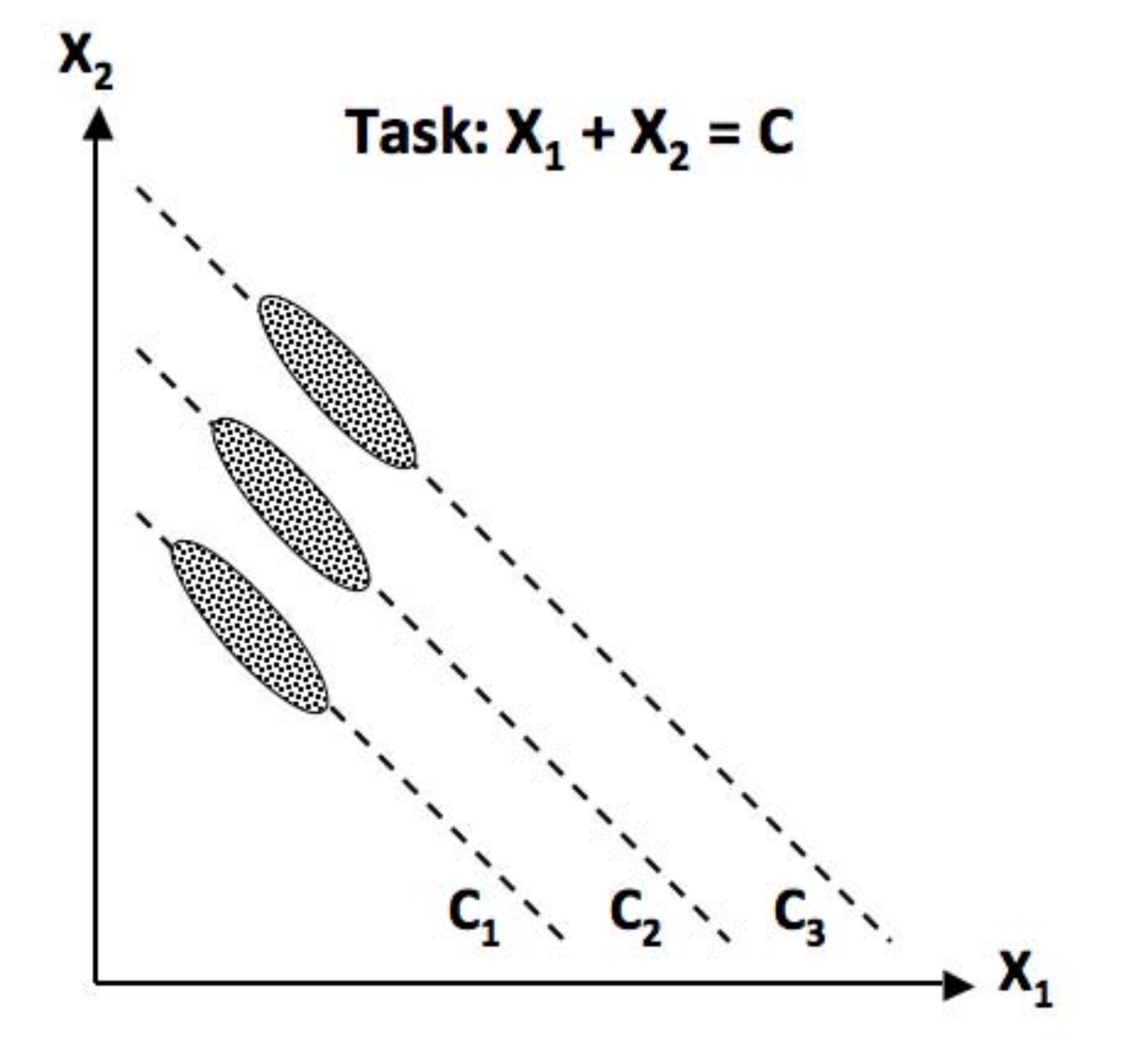}
\caption{Consider a task of producing a constant sum of two variables, e.g.,
pressing with two fingers and producing a constant total force level. The
dashed lines show solution spaces (uncontrolled manifolds, UCMs) for three
different force levels, $C_{1}$, $C_{2}$, and $C_{3}$. Across repetitive
trials, clouds of data points form ellipses elongated along the corresponding
UCM. This shape reflect lower stability along the UCM as compared to the
orthogonal direction relevant to the task-imposed constraint. Note that the
three data clouds are centered not randomly along the UCMs but reflect a
certain preferred sharing of the task between the two effectors. This
preference may reflect an optimization principle.}
\label{FIG4a}
\end{center}
\end{figure}
Figure \ref{FIG4a} illustrates the task $x_{1}+x_{2}=C$ for different values
of $C$. While all points on the slanted dashed lines correspond to perfect
task performance, actual behavior shows much more constrained clouds of
solutions that show larger deviations along the solution space (the UCM)
compared to deviations orthogonal to that space (that lead to errors in
performance). Note that if the task is learned for a particular value of $C$,
the solutions show robustness: Similar relative locations and shapes of the
solution clouds for other values of $C$.

More formally, assuming a feedback matrix $\widehat{\beta}$ stabilizing a
$M$-dimensional subspace is found, one may think of imposing complementary
constraint, that can either be in the form of explicit equations or follow
from a requirement of minimization of a cost function. In such a setting, one
speaks about stabilization of a subspace of a dimension $M_{+}$ exceeding the
dimensionality $M$ of the initial task subspace. Such perfectionism may be
viewed as a secondary tasks decreasing variance in some directions of the UCM.

We would like to emphasize once again that there exists three types of
specific bases that can be employed for the dynamics description: (1)A
laboratory basis of measured variables, (2) A basis of modes resulting from
both body- and task-imposed constraints, while each mode has independent noise
statistics, and (3) A task-specific basis of variables along which the control
is local. The last basis is a conceptually new feature affording a simple
structure of the control algorithm.

The relation between the second and the third bases may change during the
process of stability search: Application of the local control $\widehat{G}$
leads to a change of $\widehat{\omega}=\widehat{\alpha}\widehat{G}%
\widehat{\beta}$, that is, the way the task requirement $\widehat{\alpha}$ is
mapped onto the feedback action. Thereby it may affect both the basis of the
noise-decorrelated modes and the magnitude of the corresponding modes
susceptibilities. There exists such an extreme, where the mode
susceptibilities $\left\vert \mathcal{S}_{i}-i\omega\right\vert ^{-2}$ vary
without changing the corresponding eigenvectors $\overrightarrow
{\mathcal{X}}_{i}$ of $\widehat{\Omega}$ -- the linear combinations
$\overrightarrow{\mathcal{X}}_{i}$ and $\overrightarrow{\mathcal{X}}_{j}$ of
the laboratory basis variables remain statistically independent, and only the
variance $\left\langle \delta x_{i}^{2}\right\rangle $ of one combination
starts to exceed that of the other $\left\langle \delta x_{j}^{2}\right\rangle
$. One can call this regime "mode crossing", by the analogy to the phenomenon
of "term crossing" in Quantum Mechanics \cite{Akulin2014}. In the general
case where, along with a change of the susceptibilities, the local changes
equally result in the emergence of an appreciable covariance between the modes
$\left\langle \delta x_{i}\delta x_{j}\right\rangle \neq0$, one encounters the
phenomenon of so-called "avoid crossing", where formerly the larger mode
susceptibility, though approaching in magnitude the other one, remains always
larger, while both eigenvectors $\overrightarrow{\mathcal{X}}_{i}$ and
$\overrightarrow{\mathcal{X}}_{j}$ corresponding to these modes rotate.

\subsection{Algorithm for Hierarchical Feedback}

Hierarchical control considered in this section gives an example of 
stabilization search algorithm different from that considered earlier. We
demonstrate that such an algorithm requires only local control, whereas the
role of nonlocal control can be played by another randomly chosen linear
feedback matrix once the current one does not yield stabilization. The idea of
hierarchical control of the human body is very old. Arguably, the first
hierarchical system considered the brain and the spinal cord. A comprehensive
scheme of control with referent configurations has been suggested recently
built on a hierarchical principle, starting with referent values for a few
task-specific, salient variables, and resulting in referent length values for
numerous involved muscles \cite{LatashML(2010)}. An example of hierarchical control is
the command structure of an army, where a general gives orders to privates
through the commanders of regiments, battalions, and platoons. We explore
efficiency of hierarchical control in this section, along with another
modification -- the most nimble $x_{i}$ is no longer punished by a step change
of the sign of its contribution but experiences a smooth change of the
participation in the feedback as the cosine of the corresponding functional
$z_{i}.$

\subsubsection{Intrinsic instability of the hierarchical control}

In mathematical terms, the time derivative of a vector $\overrightarrow{x}%
_{n}$ of variables at $n$-th step depends on the variables $\overrightarrow
{x}_{n-1}$at the previous step, while the spatial dimension $N_{n}$ of each
next step varies from step to step. At each step local control may be
implemented. For example, the corresponding set of the local control
equations for a three-level hierarchy has the form
\begin{align}
\overrightarrow{y}  &  =\widehat{\alpha}\overrightarrow{x}_{3}\label{EQ18}\\
\frac{d}{dt}\left.  x_{1}\right.  _{i}  &  =\left(  \widehat{G}_{1}\right)
_{ii}\left(  \widehat{A}\overrightarrow{y}\right)  _{i}\nonumber\\
\frac{d}{dt}\left.  x_{2}\right.  _{i}  &  =\left(  \widehat{G}_{2}\right)
_{ii}\left(  \widehat{B}\overrightarrow{x}_{1}\right)  _{i}\nonumber\\
\frac{d}{dt}\left.  x_{3}\right.  _{i}  &  =\left(  \widehat{G}_{3}\right)
_{ii}\left(  \widehat{C}\overrightarrow{x}_{2}\right)  _{i},\nonumber
\end{align}
based on the diagonal matrix elements \\
$\left(  \widehat{G}_{n}\right)_{ii}=\cos\left[  \left.  z_{n}\right.  _{i}(t)\right]$
of the local control
operators $\widehat{G}_{n}$. If at each level $n$ of the hierarchy, we also
include diagonal damping matrices $\widehat{\gamma}_{n}$ of the dimension
given by the number $N_{n}$ of the variables at this level, the set of
equations (\ref{EQ18}) can be seen as a single matrix equation

\begin{multline}
\left(
\begin{array}[c]{c}
\left(  \frac{d}{dt}-\widehat{\gamma}_{1}\right)  \overrightarrow{x}_{1}\\
\left(  \frac{d}{dt}-\widehat{\gamma}_{2}\right)  \overrightarrow{x}_{2}\\
\left(  \frac{d}{dt}-\widehat{\gamma}_{3}\right)  \overrightarrow{x}_{3}%
\end{array}
\right)  =   \\
\left(
\begin{array}[c]{ccc}
\widehat{G}_{1} & 0 & 0\\
0 & \widehat{G}_{2} & 0\\
0 & 0 & \widehat{G}_{3}%
\end{array}
\right)  \left(
\begin{array}
[c]{ccc}%
0 & 0 & \widehat{A}\widehat{\alpha}\\
\widehat{B} & 0 & 0\\
0 & \widehat{C} & 0
\end{array}
\right)  \left(
\begin{array}[c]{c}
\overrightarrow{x}_{1}\\
\overrightarrow{x}_{2}\\
\overrightarrow{x}_{3}
\end{array}
\right) .
\label{EQ19}
\end{multline}

Alternatively, the local control operation may be applied not to
susceptibility of a given variable to external factors, but to efficiency with
which this variable acts on the variables of the next generation. In such a
case, two operators on the left hand side of Eq.(\ref{EQ19}) have to be
interchanged
\begin{multline}
\left(
\begin{array}[c]{c}
\left(  \frac{d}{dt}-\widehat{\gamma}_{1}\right)  \overrightarrow{x}_{1}\\
\left(  \frac{d}{dt}-\widehat{\gamma}_{2}\right)  \overrightarrow{x}_{2}\\
\left(  \frac{d}{dt}-\widehat{\gamma}_{3}\right)  \overrightarrow{x}_{3}%
\end{array}
\right)  = \\
\left(
\begin{array}[c]{ccc}
0 & 0 & \widehat{A}\widehat{\alpha}\\
\widehat{B} & 0 & 0\\
0 & \widehat{C} & 0
\end{array}
\right)  \left(
\begin{array}[c]{ccc}
\widehat{G}_{1} & 0 & 0\\
0 & \widehat{G}_{2} & 0\\
0 & 0 & \widehat{G}_{3}
\end{array}
\right)  \left(
\begin{array}[c]{c}
\overrightarrow{x}_{1}\\
\overrightarrow{x}_{2}\\
\overrightarrow{x}_{3}
\end{array}
\right) .
\label{EQ19a}
\end{multline}

Though offering a simple way to address many variables at once, hierarchical
control may add instability. Therefore, the intermediate steps have
to be damped in order to avoid such an instability in contrast to one-step
control of Eq.(\ref{EQ9}). In particular, note that for zero damping
$\widehat{\gamma}_{i}=0$, even the simplest two-level control becomes
unstable, and this is always the case for a higher number $l$ of the control
levels. The root of this instability is rather simple, and can be illustrated
with an example of a three-level control scheme, with just one variable at
each level, when Eq.(\ref{EQ19}) takes the form
\begin{equation}
\begin{array}[c]{c}
\frac{d}{dt}x_{1}=G_{1}A\alpha x_{3}\\
\frac{d}{dt}x_{2}=G_{2}Bx_{1}\\
\frac{d}{dt}x_{3}=G_{3}Cx_{2}
\end{array}
\label{EQ20}
\end{equation}
corresponding to the characteristic equation
\begin{equation}
\lambda^{3}=G_{3}CG_{2}BG_{1}A\alpha .
\label{EQ21}
\end{equation} 
It is evident that, whatever a non-zero complex number standing on the right
hand side of this equation is, the phase factors of the three roots of the
cubic equation are equally distributed on the unit circle in the complex plane
such that at least one of them has a positive real part. The same structure of
the root distribution persists in a higher-dimensional case with $l$-level
control, since the characteristic equation in this case has the form
\begin{equation}
\mathrm{Det}\left\vert \lambda^{l}-\widehat{G}_{l}\ldots\widehat{G}
_{3}\widehat{C}\widehat{G}_{2}\widehat{B}\widehat{G}_{1}\widehat{A}
\widehat{\alpha}\right\vert =0 ,
\label{EQ22}
\end{equation}
and hence the roots of the characteristic polynomial given by $l$-th roots of
the eigenvalues of the matrix $\widehat{G}_{l}\ldots\widehat{G}_{3}\widehat
{C}\widehat{G}_{2}\widehat{B}\widehat{G}_{1}\widehat{A}\widehat{\alpha}$ are
also uniformly distributed on the corresponding circles in the complex plane,
with radii given by the eigenvalues moduli.

We thus come to the conclusion that damping is indispensable for stable
hierarchical control. At the same time, all $\widehat{\gamma}_{i}>0$ imply a
trivial asymptotic situation of completely damped motion at all levels for
all tasks. Therefore, in order to have a reasonable model, we have to assume
vanishing of the damping rates at only one hierarchical level of control.
Presumably, this level should have the maximum dimensionality $N_{n}$.

\subsubsection{Action of the feedback loops}

\begin{figure}[ht]
\begin{center}
\includegraphics[width=3.6in]{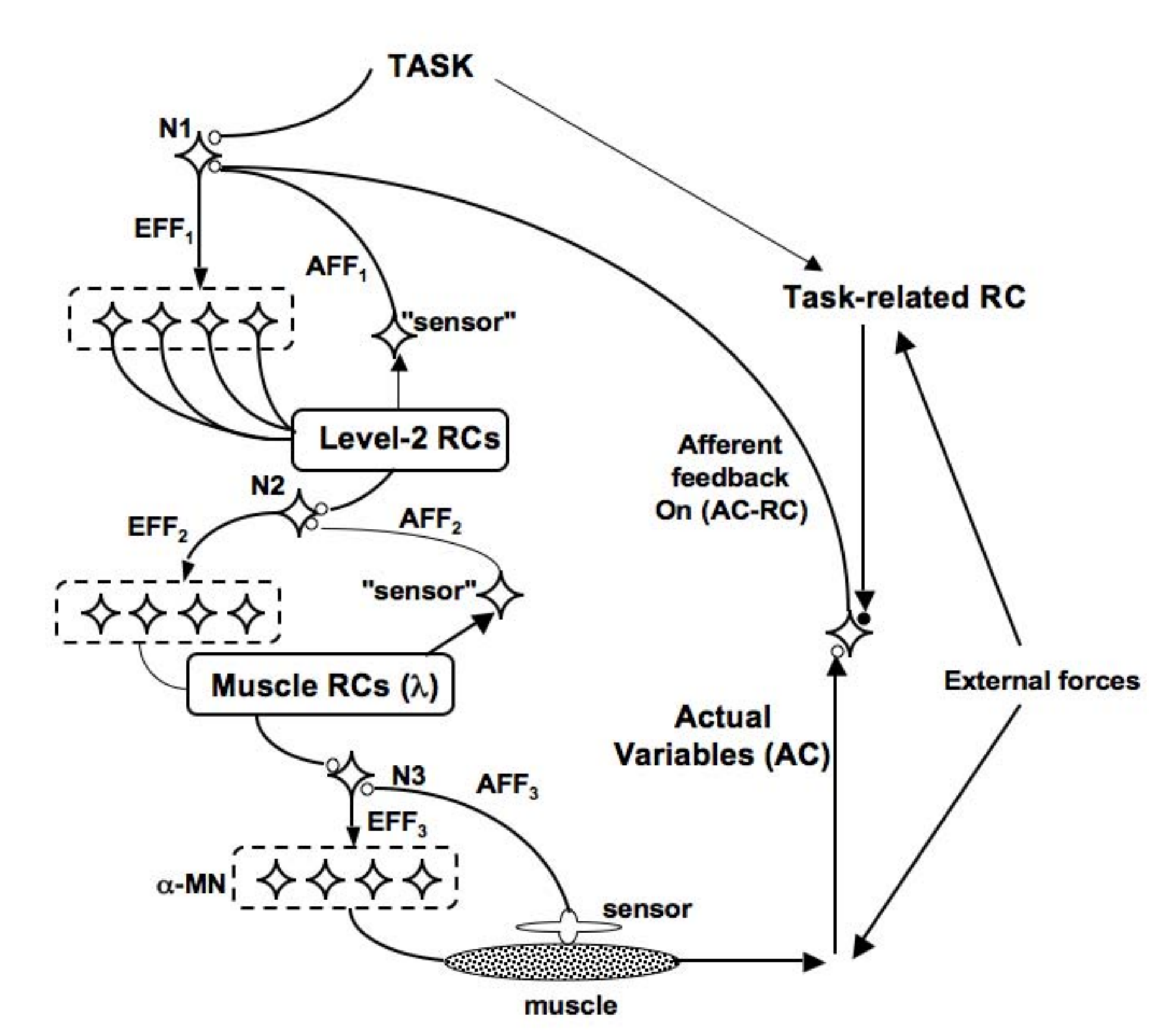}
\caption{A scheme of hierarchical control of the hand within the idea of
control with referent configurations (RCs) of the body\cite{FeldmanAG(2009)}.
At the top level, a low-dimensional set of
referent values for salient, task-specific variables is reflected in the RC. A
sequence of few-to-many transformations results in higher-dimensional RCs at
the digit level and muscle level. Local feedback loops ensure stability with
respect to the variables specified by the input. The global feedback loop
ensures that the actual body configuration moves towards one of the solutions
compatible with the task RC. At each level, inputs to a neuronal pool ($N1$,$N2$,
and $N3$) are combined with afferent feedback (AFF) to produce the output
(efferent signals, EFF). At the lowest level, elements are alpha-motoneurons
and their referent coordinates correspond to the thresholds of the tonic
stretch reflex (lambda). Modified by permission from \cite{LatashML(2010)}.}
\label{Fig6t}
\end{center}
\end{figure}

We note that stability can be to a certain extend improved by introducing
intermediate feedback loops to the net feedback loop as shown in
Fig.\ref{Fig6t}.

For the example of aforementioned three-level hierarchical control, the
Laplace-transformed equations of motion for the case of feedback loop equipped
with secondary feedback loops read
\begin{align}
\lambda\overrightarrow{x}_{1} &  =\widehat{G}_{1}\widehat{A}\widehat{\alpha
}\overrightarrow{x}_{3}+\widehat{\Lambda}_{2}\overrightarrow{x}_{2}\nonumber\\
\lambda\overrightarrow{x}_{2} &  =\widehat{G}_{2}\widehat{B}\overrightarrow
{x}_{1}+\widehat{\Lambda}_{3}\overrightarrow{x}_{3}\nonumber\\
\lambda\overrightarrow{x}_{3} &  =\widehat{G}_{3}\widehat{C}\overrightarrow
{x}_{2} 
\label{EQ22a}
\end{align}
where the secondary feedback loops of the second and the third hierarchical
levels are given by the matrices $\widehat{\Lambda}_{2}$ and $\widehat
{\Lambda}_{3}$, respectively. The latter can be also considered as dependent
on local variables $z$. For a given set of variables $z$, the
instantaneous eigenvalues $\lambda$ can be calculated as the roots of an
analog of Eq.(\ref{EQ22}), which reads
\[
\mathrm{Det}\left\vert \lambda-\widehat{G}_{3}\widehat{C}\frac{1}
{\lambda-\widehat{G}_{2}\widehat{B}\widehat{\Lambda}_{3}}\widehat{G}
_{2}\widehat{B}\frac{1}{\lambda-\widehat{G}_{1}\widehat{A}\widehat{\alpha
}\widehat{\Lambda}_{2}}\widehat{G}_{1}\widehat{A}\widehat{\alpha}\right\vert
=0 .
\]

Note that in the basis where the matrices $\widehat{G}_{2}\widehat{B}
\widehat{\Lambda}_{3}$ and $\widehat{G}_{1}\widehat{A}\widehat{\alpha}
\widehat{\Lambda}_{2}$ are diagonal, they can be interpreted as damping
matrices $\widehat{\gamma}_{1}$ and $\widehat{\gamma}_{2}$, respectively,
although in this case these matrices are dependent on parameters $z$, which,
generally speaking, in this representation cannot be considered as local
control. Still, this brings about an idea of the self-adjusting damping
dependent on time via the parameters $z_{i}(t)$.

\subsubsection{Stabilization by self-adjusting damping}

Adjusting gain in the feedback loops controls stability at each hierarchical
level. This effect can be modeled when the damping parameters $\gamma_{i}$ are
taken depending on the local parameters $z_{i}$, increasing, for instance, as
\begin{equation}
\gamma_{i}=\Gamma_{i}+\mu_{i}z_{i}
\label{EQ22b}
\end{equation} ,
for the variables at each damped level. The parameters $\gamma_{i}$ are
positive for all but one hierarchical levels, where they may remain zero, in
order to avoid the trivial case of complete damping of all variables. An
analog of the characteristic equation (\ref{EQ22}) for this case gets
simplified and reads
\[
\mathrm{Det}\left\vert \lambda-\widehat{\gamma}_{3}-\widehat{G}_{3}\widehat
{C}\frac{1}{\lambda-\widehat{\gamma}_{2}}\widehat{G}_{2}\widehat{B}\frac
{1}{\lambda-\widehat{\gamma}_{1}}\widehat{G}_{1}\widehat{A}\widehat{\alpha
}\right\vert =0,
\]
where one of the three diagonal matrices $\widehat{\gamma}_{1}$,
$\widehat{\gamma}_{2}$ or $\widehat{\gamma}_{3}$ is assumed to be zero.

In turn, for the local control over the damping rate, the dependence
$z_{i}(t)$ can be given by a differential equation relating the positive rate
of increase $\frac{d}{dt}z_{i}$ with an even power $f$ of either the
corresponding variable,%
\begin{equation}
\frac{d}{dt}z_{i}=\epsilon\left(  x_{i}\right)  ^{f}, \label{EQ23o}%
\end{equation}
or the corresponding variable velocity,
\begin{equation}
\frac{d}{dt}z_{i}=\epsilon\left(  \frac{dx_{i}}{dt}\right)  ^{f}, \label{EQ23}%
\end{equation}
where $\epsilon<1$ is a numerical parameter, which may depend on the hierarchy
level number and the variable number.

In the following Table \ref{table3}
\begin{table*}[ht]
\[
\begin{array}[c]{|c|cccccc|}
\hline
M & 1 & 2 & 3 & 4 & 5 & 6\\
N_{1}/N_{2}/N_{3} & 5/50/200 & 5/50/200 & 5/50/200 & 5/50/200 & 5/50/200 &
6/50/200\\
R & 1 & 1 & 0.69 & 0.43 & 0.3 & 0.15 \\
\hline
\end{array}
\]
\caption{The stabilization success rate of a $M$-dimensional subspace for 
 hierarchical control with $N_1/N_2/N_3$-dimensional levels. }
\label{table3}
\end{table*}
we present the success rates calculated for series of $100$ random casts of
the rectangular feedback matrices $\widehat{A}$, $\widehat{B}$, and
$\widehat{C}$ corresponding to the dimensions $N_{1}$, $N_{2}$, and $N_{3}$ of
the spaces of $x_{1}$, $x_{2}$, and $x_{3}$, respectively, and generated
independently each time for a random cast of the $N_{3}\times M$ task matrix
$\widehat{\alpha}$.  The identical damping parameters $\Gamma_{i}=0.1$ and
$\mu_{i}=1$ have been chosen. Moreover, in order to avoid the trivial
situation yielding $100\%$ success rate, where all variables get strongly damped
and hence become vanishing in the course of time, a fourth hierarchical level
corresponding to the vector variables $\overrightarrow{x}_{0}$ defined by the
differential equation without damping, $\frac{d}{dt}\overrightarrow{x}%
_{0}=\overrightarrow{y}$, has been included into Eq.(\ref{EQ20}). The non-zero
diagonal matrix elements of the local feedback control matrices $\widehat
{G}_{1,2,3}$ have been chosen here as $G_{ii}=\cos\left( 0.1z_{i}\right) $,
and the power index $f$ and the numerical parameter $\epsilon$ in
Eq.(\ref{EQ23}) have been chosen equal to $4$, and $0.1$, respectively for all variables.

\begin{figure}[ht]
\begin{center}
\includegraphics[width=3in]{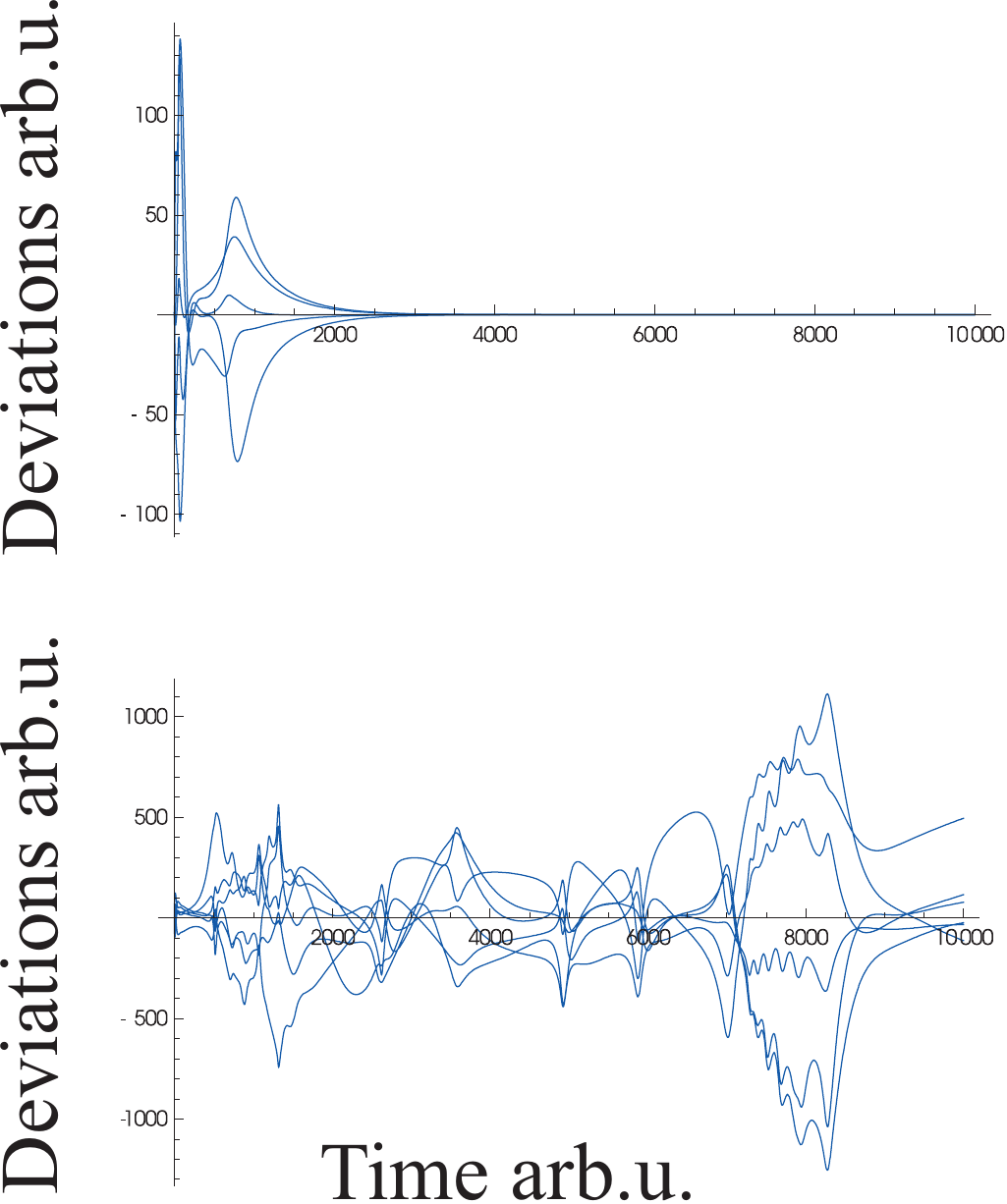}
\caption{Time dependencies for the successfully controlled five-dimensional
subspace in the hierarchical cascade setting with $N_{1}/N_{2}/N_{3}=5/50/200$
(upper). Unsuccessful control (lower).}
\label{FIG6}
\end{center}
\end{figure}

A typical time dependence $\overrightarrow{y}(t)$ for the successfully
stabilized subspace is shown in Fig.\ref{FIG6} (the top plot).
In accordance with the smoothness of the local feedback control dependence
($\cos\left( 0.1z_{i}\right)  $ instead of $\mathrm{sign}(10\pi-z_{i})$), the
time dependence does not manifest a cusp-like changes of the derivatives,
typical of those seen in Fig.\ref{FIG1}. For unsuccessful control, the
dependence looks like the one depicted in Fig.\ref{FIG6} (the bottom plot).

A rather high success rate of the hierarchical local control over the dynamics
for random choice of the feedback matrices and self-adjusting damping suggests
a strategy, which can replace the non-local control. Once the current random
cast of the feedback matrices does not yield stabilization during a run, one
may achieve it by taking another random cast in the course of the same run,
and in case of failure, repeat the attempt again and again. In particular, the
probability $0.3$ implies that just a few ($4-5$) such casts during a run are
required in order to stabilize a $5$-dimensional subspace. Changing the entire
linear feedback matrices when the local control turns out to be inefficient
may be considered as an action replacing nonlocal control discussed in the
previous section.

\begin{figure*}[ht]
\begin{center}
\includegraphics[height=2.4in]{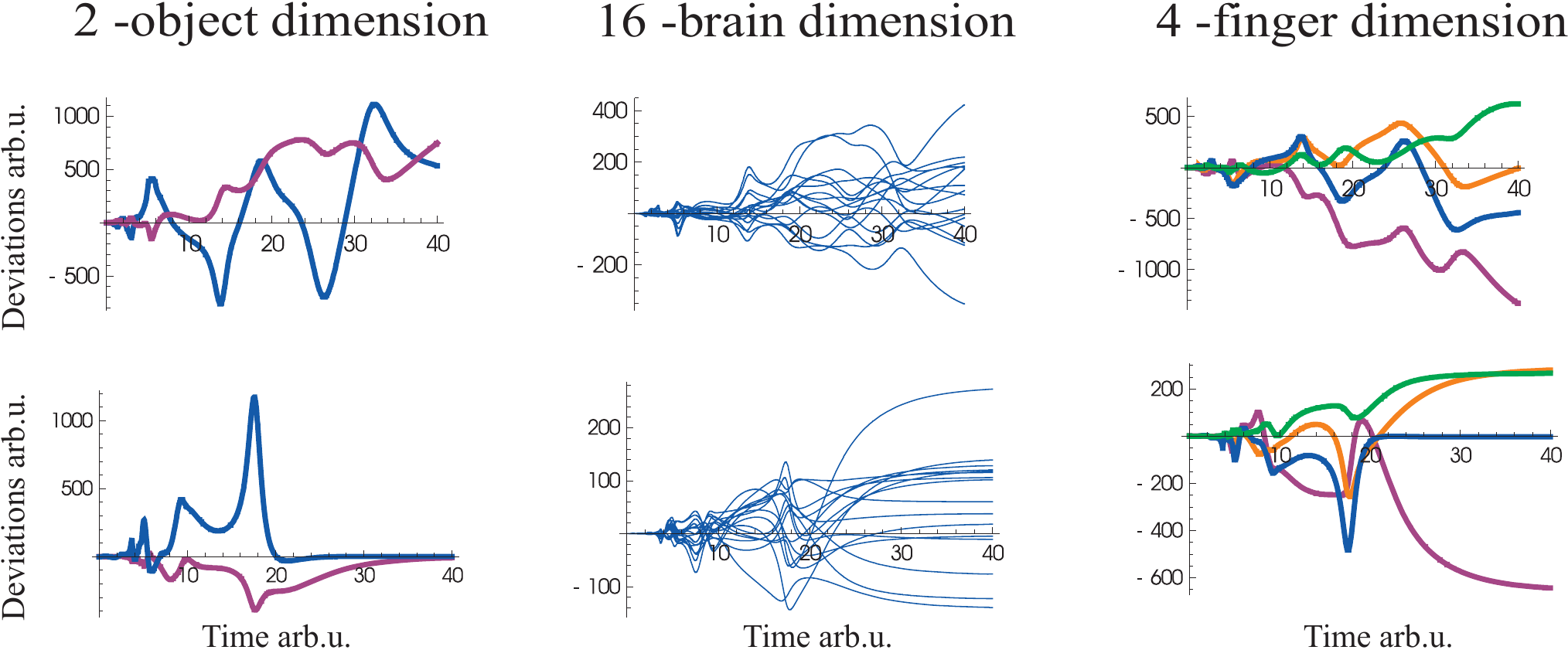}
\caption{Examples of successful (lower panels) and unsuccessful (upper panels)
search for the new equilibrium control. We show coordinate variables as
functions of time in arbitrary units. First column: deviations of two controlled coordinates from their prescribed
value -- zero. The second column: dynamics of the hypothetical elemental variables tend to
some asymptotic values for the successful control. Third column: dynamics of
motor variables. One sees the dynamics of transition from the old equilibrium
position to the new one.}
\label{FIG8}
\end{center}
\end{figure*}

Note, that the dimensionality at each step of the hierarchy does not need to
be larger than the dimensionality of the previous step. We illustrate this by
an example of control over a two-dimensional space $y$ exerted by a three-step
hierarchical feedback with locally changing parameters. The first-step variables
$\overrightarrow{x}_{1}$ belong to a space of the dimension $N_{1}=4$; here
the damping increases according to Eqs.(\ref{EQ22b},\ref{EQ23}) with
$\epsilon=0.2$, $f=2$, and $\mu_{i}=0.1$. Next step of the hierarchy comprises
the variables $\overrightarrow{x}_{2}$ that belong to a space of the dimension
$N_{2}=16$ where they experience no damping. The last step of the hierarchic
feedback comprises variables $\overrightarrow{x}_{3}$ in a four-dimensional
space, $N_{3}=4$, with strong constant damping $\Gamma_{i}=5$ at the third
level. The success rate for these parameters was $R\simeq0.8$. Examples of
unsuccessful and successful search of stability are shown in Fig.\ref{FIG8}
for the components of the vectors $y$, $\overrightarrow{x}_{2}$, and
$\overrightarrow{x}_{3}$.
One sees that for successful control the components of the high-dimensional
vectors tend to asymptote with time, while for unsuccessful control, they
keep changing.

\section{Experiments with Human Hand}

An experiment was performed to illustrate one of the central ideas of the
suggested scheme and check some of its predictions. We used the task of
accurate force and moment-of-force production by the four fingers of the
dominant hand. The forces produced by four fingers, the index, the middle, the
ring, and the little, were controlling the position of a point on the computer
screen. Two types of perturbations were used. First, we modified the visual
feedback leading to changes in the mapping between the finger forces and the
two task variables. Second, we used mechanical perturbations applied to a
finger that led to actual changes in the two task variables. The experiments
were approved by the Office for Research Protections at the Pennsylvania State University.

\subsection{Visual Feedback Perturbation}

The first experiment was as follows. Four $6$-axis force/moment sensors
(Nano-$17$, ATI Industrial Automation, USA) mounted on the table were used to
measure normal forces produced by the tips of the index ($I$), middle ($M$),
ring ($R$), and little ($L$) fingers. To increase friction between the digits
and the sensors, $320$-grit sandpaper (SandBlaster,$3M$, USA) was
placed on the contact surfaces of the sensors. The centers of the sensors were
evenly spaced at $30\;mm$. The output analog signals from the sensors where
digitized with the $16$-bit data acquisition card (PCI-6225, National
Instrument, Austin, TX, USA) at $100\;Hz$. A LabVIEW program (LabVIEW 2011,
National Instrument, USA) was used to provide visual feedback and store the
data on the computer (Dell Inc., USA). Offline processing and analysis was
done in Matlab (MathWorks, USA).

At the beginning, the four-dimensional space of the finger forces was mapped
onto the two-dimensional space of the screen according to a very natural rule
- the vertical coordinate was proportional to the net resultant finger force,
while the horizontal displacement was given by the net moment of the finger
forces computed with respect to a horizontal axis passing in the
anterio-posterior direction in-between the R and M finger sensors. The subject
was first requested to place the cursor into a position in the middle of the
screen and, second, to keep it in this position at all times. Once this task
was accomplished, the law according to which the deviation of the finger
forces from the steady-state finger force values are mapped to the deviation
of the point from the center of the screen was changed without the subject's
knowledge. The new law relating the four-dimensional space of the force
deviations with the two-dimensional space of the cursor deviations from the
center point was given by a new, randomly chosen, $2\times4$ Jacobian
matrix. This law drew the system to an unstable regime. In order to keep the
point at the center of the screen, the subject had to find a new feedback rule
stabilizing this position. Trajectory of the forces exerted by the fingers in
the course of this search for new stability were recorded.

\subsubsection{Results - manifestations of the feedback changes}

\begin{figure}[ht]
\begin{center}
\includegraphics[width=3in]{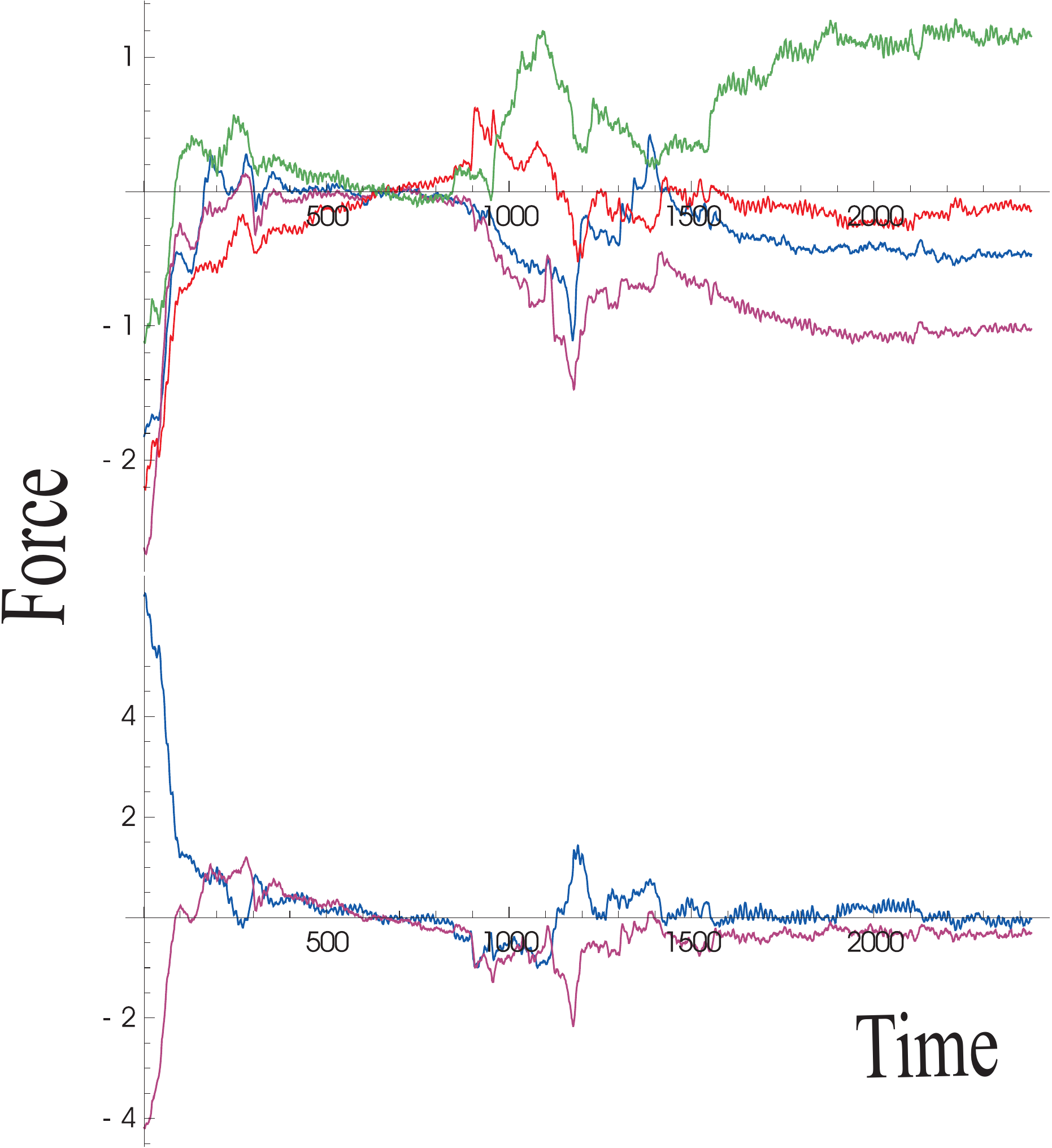}
\caption{Finger forces in $N$ as functions of time in $10^{-2}s$ (upper panel)
 and the coordinates of the point position on the screen (lower panel). The
law relating the forces to the position of the point was changed at $t=7s$,
and the values of the finger forces at that time were chosen as zero $N$.}
\label{FIGEX1}
\end{center}
\end{figure}

In Fig.\ref{FIGEX1} we present an example of the time profiles of the finger
forces (upper panel), along with the coordinates of the point on the screen
(lower panel), which were recorded during a successful trial. Overall
stabilization success rate was $R\simeq0.55$.

There is a qualitative similarity between the dependencies depicted in
Fig.\ref{FIGEX1} and the corresponding calculated dependencies illustrated in
Fig.\ref{FIG8}.

The sharp spikes and jumps on the experimental curves correspond to changes of
the trends; these events show approximately the moments of time when a
correcting action was taken similarly to the simulated curve in Fig.\ref{FIG1}
. Though spikes in the simulated curves may be unambiguously associated with
the correcting actions of the feedback sign change, within the experimental
finger force curves analysis there is no formal rules allowing to identify
such moments, and one can speak only about an intuitive similarity between the
dependencies. Moreover, since we are exploring the regime of searching for an
equilibrium in new, formerly unknown, conditions, we cannot invoke the
powerful tool of statistical analysis comparing different trials, since each
new trial corresponds to new initial and task conditions, whereas multiple
sequential repetitions of the same experimental setting for the same subject
would likely involve processes of learning and adaptation that are beyond the
scope of questions addressed here. Still certain information about the
moments of changes of the feedback matrix can be extracted from the analysis
of noises, since, as discussed in Sect.\ref{NOISE}, the
principle axes of the tensor susceptibility to noise (so-called principle
components) coincide with eigenvectors of the dynamic matrix $\widehat{\Omega
}$, and when directions of these axes change, the matrix $\widehat{\Omega}$
changes as well, thus implying a change of the feedback matrix. Moreover, the
larger the eigenvalue $\mathcal{C}_{i}$ of $\widehat{\Omega}$ is, the smaller
is the susceptibility of the corresponding direction to noise, in accordance
with Eq.(\ref{EQ13}). The noise analysis is discussed in the next section.

\subsubsection{Analysis of the principal components of noise covariance
reveals feedback changes}

Human fingers are not independent force generators: When a person tries to
press with one finger, other fingers of the hand show unintentional force
increase \cite{KilbreathSL,LiZM}. This phenomenon has
been addressed as enslaving or lack of finger 
individuation \cite{ZatsiorskyVM(2000),SchieberMH}. Patterns of enslaving are person-specific
and relatively robust; changes in these patterns have been reported with
specialized practice \cite{SlobounovS, WuYH}. These
patterns may be described as eigenvectors of enslaving $\overrightarrow
{f}_{E,i}=\left\{  f_{j}\right\}  _{E,i}$, where $i=\left\{  I-\mathrm{index}%
;M-\mathrm{middle},R-\mathrm{ring},\right.  $ and$\left.  L-\mathrm{little}%
\right\}  $ stands for an instructed finger. Directions of $\overrightarrow
{f}_{E,i}$ may be viewed as preferred directions in the space of finger forces
when the person is trying to press with individual fingers They are related
as
\begin{equation}
\overrightarrow{f}_{E,i}=\widehat{U}\overrightarrow{X}_{i} \label{EQ23a}%
\end{equation}
 to the forces $X_{i}$ of the individual fingers $i$ by an orthogonal matrix
$\widehat{U}$ representing rotation in the four-dimensional space. We assume,
that these very directions may change during the search of stability and they
manifest statistically independent fluctuations thus being the eigenvectors
$\overrightarrow{\mathcal{X}}_{i}$ of the deviation covariance matrix.

Since in our experiment the task was formulated in a two-dimensional space, and
hence the rank of the dynamic problem presumably equals two, there should be
only a two-dimensional sub-space in the space of the finger forces that governs
the dynamics of the point on the screen. This implies that
only two out of four eigenvalues of the matrix $\widehat{\Omega}$ differ from
zero, and the other two vanish, unless an external to the task requirement
associated with additional constraints is imposed upon the system, as mentioned at the end of Sect.\ref{NOISE}. In the latter case, all the
eigenvalues do differ from zero, but they are decreasing exponentially with
each next eigenvalue being scaled, on average, by a factor, as it was the case
for the hierarchy of nimbleness shown in Fig.\ref{FIG2}. This has a
consequence important for the noise analysis: The directions that belong to
the two-dimensional sub-space relevant to feedback are less susceptible to
noise, while the noisy directions correspond to the null-space and do
not contribute significantly to the feedback.

The dependence $X_{i}(n)$ of the forces produced by the I, M, R, and L fingers
was captured at the sequential time points $t=n\tau$ separated by time
intervals of $\tau=10^{-2}\;sec$. The covariance matrix was extracted from
these experimental data in several steps. First, for each $X_{i}(n)$ an
average time dependence $x_{i}(n)=x_{i}(t)$ has been calculated numerically as
\begin{equation}
x_{i}(t)=\sum_{n}\frac{X_{i}(n)}{Y\sqrt{\pi}}\exp\left[  -\frac{\left(
t-n\right)  ^{2}}{Y^{2}}\right] , 
\label{EQ24}
\end{equation}
where $\tau$ is chosen as a time unit, while the averaging is performed over a
time window $\sim Y\tau$ with a width $Y$. Next, the covariance of
the finger forces
\begin{equation}
C_{i,j}(t)=\sum_{n}\frac{\delta x_{i}(n)\delta x_{j}(n)}{Y\sqrt{\pi}}
\exp\left[  -\frac{\left(  t-n\right)  ^{2}}{Y^{2}}\right]  , 
\label{EQ25}
\end{equation}
where
\[
\delta x_{i}(n)=X_{i}(n)-x_{i}(n),
\]
was found numerically with the same Gaussian width $Y$.

Being a real and symmetric matrix, $C_{i,j}$ can be set in a diagonal form by
an orthogonal transformation given by a rotation matrix $U_{i,j}(t)$ and its
inverse matrix $U_{i,j}^{-1}(t)$, such that
\begin{equation}
C_{i,j}(t)=\sum_{k}U_{i,k}(t)\mathcal{C}_{k}(t)U_{k,j}^{-1}(t). 
\label{EQ26}
\end{equation}
The eigenvalues $\mathcal{C}_{k}(t)$ provide us with the principle components
of the noise in the orthogonal directions of statistically-independent modes,
while the matrix $U_{i,k}(t)$, which in the laboratory basis can be viewed as
a row of the column eigenvectors $\overrightarrow{\mathcal{X}}_{i}(t)$,
relates these modes to the individual finger variables. All these quantities
were found numerically from the data obtained for $C_{i,j}(t)$. Note that thus
obtained orthogonal matrix $U_{i,k}(t)$ experiences a time evolution
corresponding to rotation in the $4$-dimensional space of the finger forces,
while the angular velocity of this rotation can be found as the eigenvalues of
the left logarithmic derivative of $U_{i,k}(t)$ defined as
\begin{equation}
\widehat{R}(t)=\frac{1}{i}\frac{\partial\widehat{U}}{\partial t}\widehat
{U}^{-1}. 
\label{EQ27}
\end{equation}
The eigenvalues $R_{i}$ of $\widehat{R}(t)$ are real and have pairwise
opposite signs, such that only two real numbers characterize rotation in the
four-dimensional space. We calculate these quantities replacing the derivative
in Eq.(\ref{EQ27}) with the finite difference between two neighboring integer time
points. In order to get rid of the so-called shot noise, which is an
error-inducing influence of such a replacement, the calculation must be
followed by averaging over a time interval shorter than $Y$.

\begin{figure}[ht]
\begin{center}
\includegraphics[width=2.6in]{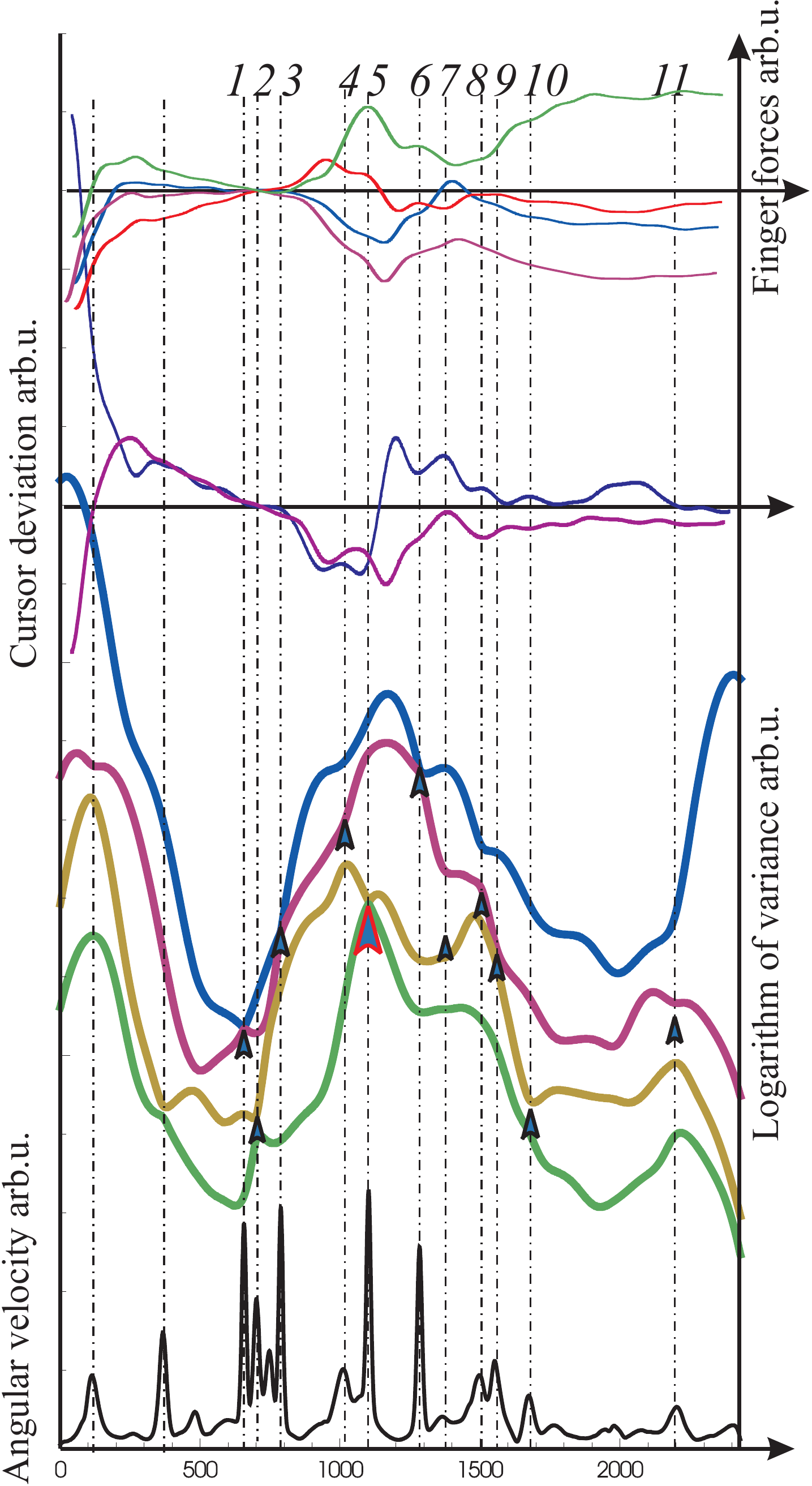}
\caption{The angular velocity $R_{1}(t)$ of rotation of the basis of the
covariance matrix eigenvectors as function of time (bottom, solid line,
arbitrary units). Logarithms of the eigenvalues $\mathcal{C}_{k}(t)$ of the
covariance matrix (four bold curves above the velocity). The average of the
covariance matrix was computed over $Y=100$ sequential time points with the
interval of $10^{-2}\;sec$ between the points. The switching of the feedback
eventually took place at the domains of "avoid crossings" discussed at the end
of Sect.\ref{NOISE}, marked here by arrows, and corresponding to the maxima of
the angular velocity. On the top of the plots, the finger forces and the
cursor coordinates, corresponding to the dependencies in Fig.\ref{FIGEX1}
averaged over the same time intervals $Y=100$, are shown for comparison. The
 "avoid crossing" of the three eigenvalues occur for the 4-th, 7-th, and 8-th
intersections. The strongest contribution comes from the 5-th crossing, which
presumably is relevant to changing of the direction in the two-dimensional
orthogonal subspace.}
\label{FIGEX2}
\end{center}
\end{figure}

Results of such processing of the experimental data presented in
Fig.\ref{FIGEX1} are shown in Fig.\ref{FIGEX2}.

One can identify eleven rotations of the covariance matrix basis
$\overrightarrow{\mathcal{X}}_{i}(t)$ presumably associated with changes of
the feedback matrix. Note that the highest rotation velocity does not
necessarily produce a strong effect on the finger forces, since the relevant
quantity rather corresponds to the spike area, representing the rotation
angle. The situation has much in common with dynamics of so-called adiabatic
and diabatic molecular term crossings, well-known in Quantum Mechanics\cite{Akulin2014}. The maximum rotation velocity corresponds to the time
moments when two or several eigenvalues of the matrix have a tendency to
coincide, thus getting rid of the difference between the large noise typical
of the null-space and small noise typical of the orthogonal subspace in a
stationary regime.

\subsection{Experiments with Mechanical Perturbations}

The results of the first experiment have demonstrated qualitative
consistency of the model and the experiment. Still the main assumption of the
suggested stable control search algorithm, the principle \textquotedblleft act
on the most nimble one\textquotedblright\ (AMN) requires additional arguments.
The second experiment was designed to test this very principle. Mechanical
perturbations (lifting and lowering a finger) were applied during the
performance of an accurate multi-finger steady-state task. According to our
scheme, quick reactions to these perturbations are based on the  AMN rule. We
checked this prediction by comparing the directions of changes in the finger
force space produced by unexpected perturbations of the steady-state force
patterns (described as a vector $\overrightarrow{f}_{P,i}$) with the first
identifiable correction produced by the subjects (a correction vector,
$\overrightarrow{f}_{C,i}$). We expected the angle between $\overrightarrow
{f}_{C,i}$ and $\overrightarrow{f}_{P,i}$ to be small, smaller than the angle
between $\overrightarrow{f}_{C,i}$ and $\overrightarrow{f}_{E,i}$, defined by
Eq.(\ref{EQ23a}).

\subsubsection{Methods - the "inverse piano" setup}

Eight young, healthy subjects took part in the experiment (four males). They
were right-handed, had no specialized hand training (such as playing musical
instruments) and no injury to the hand.

An \textquotedblleft inverse piano\textquotedblright\ apparatus was used to
record finger forces and produce perturbations. The apparatus has four force
sensors placed on posts powered by linear motors, which could induce motion of
the sensors along their vertical axes (for details see Martin et al. 2011).
Force data were collected using PCB model 208C01 single-axis piezoelectric
force transducers (PCB Piezotronics, Depew, NY). The signals from the
transducers were sent to individual PCB 484B11 signal conditioners -- one
conditioner per sensor -- and then digitized at $1\;kHz$ using a $16$-bit
National Instruments PCI-6052E analog-to-digital card (National Instruments
Corp., Austin, TX). Each sensor was mounted on a Linmot PS01-23x80 linear
actuator (Linmot, Spreitenbach, Switzerland). Each actuator could be moved
independently of the others by means of a Linmot E400-AT four-channel servo
drive. Data collection, visual feedback to the subject, and actuator control
were all managed using a single program running in a National Instruments
LabView environment. Visual feedback was provided by means of a
19\textquotedblright\ monitor placed $0.8\;m$ from the subject. The feedback
cursor (a white dot) showed on the monitor represented total
finger force along the vertical axis and total moment of force in a frontal
plane computed with respect to a horizontal axis passing through the midpoint
between the $M$ and $R$ fingers along the horizontal axis. Pronation efforts
led to leftward deviation of the feedback cursor. An initial target was placed
on the screen (a white circle) corresponding to the total force of $10\;N$ and
zero total moment.

\begin{figure}[ht]
\begin{center}
\includegraphics[width=2.6in]{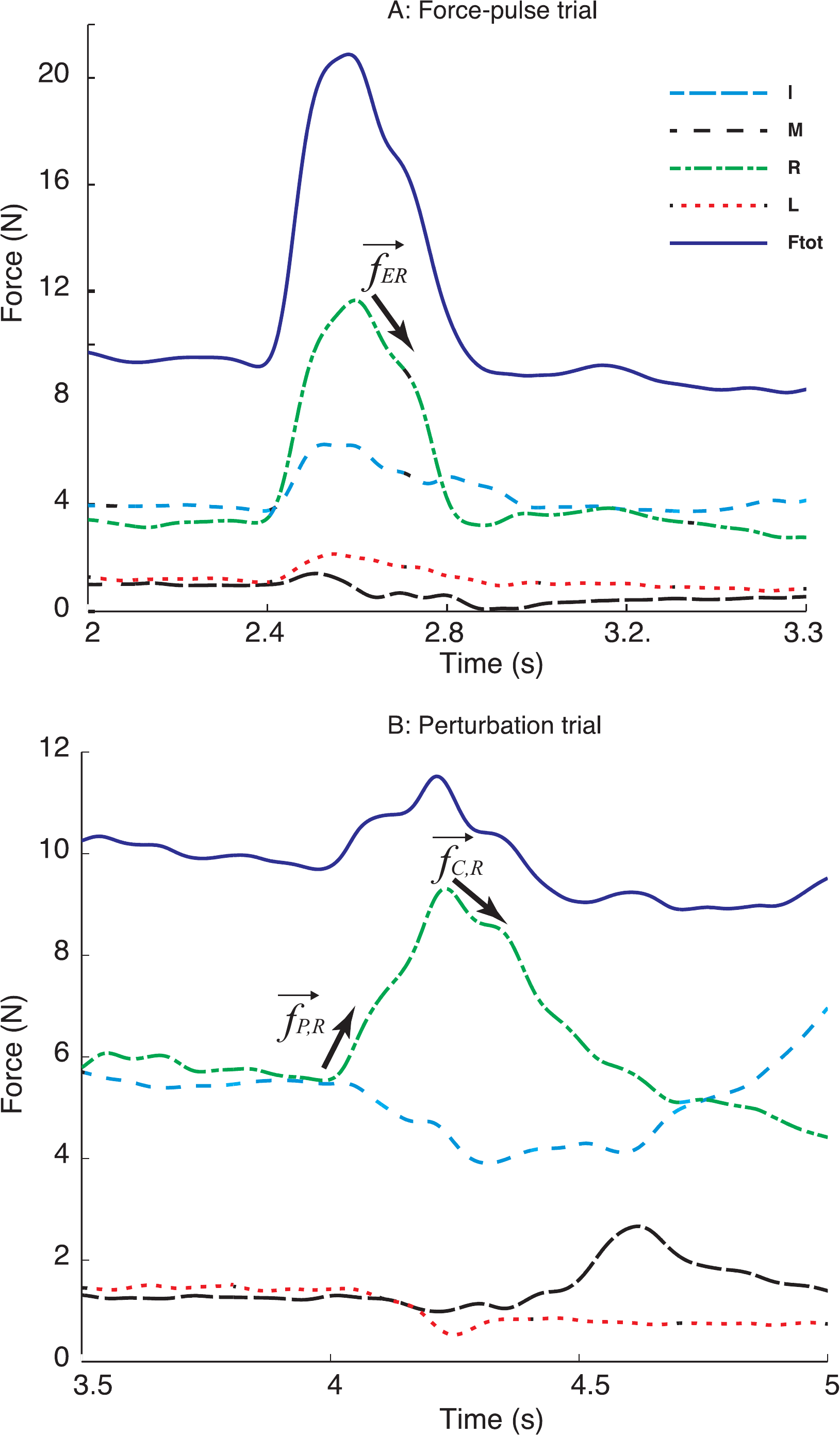}
\caption{Typical time profiles of the force-pulse trial (A, top panel) and
perturbation trial (B, bottom panel) performed by a representative subject.
The total force profile is shown with the solid trace, and individual finger
forces are shown with different dashed traces. The intervals used to compute
the eigenvectors in the force space are shown with the arrows.}
\label{FIG9}
\end{center}
\end{figure}

The experiment involved two parts: Voluntary force--pulse production and
reacting to unexpected perturbations (see later). Prior to each trial, the
subjects were asked to place the fingers on the centers of the sensors and
relax; sensor readings were set to zero during this time interval. As a
result, the sensors measured only active pressing forces. Then, a verbal
command was given to the subject and data acquisition started. The subject was
given $2\;s$ to place the cursor over the initial target. During the
force--pulse trials, the subjects were asked to produce a force pulse from the
initial target in less than $1\;s$ by an instructed finger (Figure
\ref{FIG9}A). Each finger performed three pulse trials in a random order. In
perturbation trials, one of the sensors unexpectedly moved up by $1\;cm$ over
$0.5\;s$. This led to an increase in total force (Figure \ref{FIG9}B), while
changes in the moment of force depended on what finger was perturbed. The
subject was instructed to return to the target position as quickly as
possible. Each finger was perturbed once per trial, with $10-s$ rest periods
between each of the three repetitions. Perturbation conditions were
block--randomized between fingers with $1$-min rest periods between blocks.

From the pulse trials we manually identified the onsets of the force decrease
phase (see Figure \ref{FIG9}A). From the perturbation trials we manually
identified the onsets of two time intervals: one corresponding to the
perturbation-induced force change and the other corresponding to the earliest
corrective action by the subjects. Each time interval contained $200\;ms$ of
the four--dimensional finger force ($I,M,L,$ and $R$) data. Further, for each
time interval, principal component analysis (PCA) based on the co-variation
matrix was used to compute the first eigenvector in the four-dimensional
finger force space, accounting for most variance across the time samples, for
each subject and each trial separately.

Therefore, for each perturbation trial we obtained two eigenvectors in the
finger force space. We will refer to these vectors as $\overrightarrow
{f}_{P,i}$ (force vector during the perturbation applied to the $i$-th
finger), and $\overrightarrow{f}_{C,i}$ (force vector during the earliest
correction in trials with perturbations applied to the $i$-th finger). For
each finger, from the three force--pulses trials we computed an average vector
$\overrightarrow{f}_{E,i}$ (force vector during the downward phase of force
change in the pulse task by the $i$-th finger). Note that this vector
reflected the unintentional force production by non-task fingers of the hand (enslaving).

Finally, we computed the angles $\alpha_{PE}$ (averaged across repetitions)
between $\overrightarrow{f}_{P,i}$ and $\overrightarrow{f}_{E,i}$ and angles
$\alpha_{PC}$ between $\overrightarrow{f}_{P,i}$ and $\overrightarrow{f}%
_{C,i}$ for each finger and each subject separately. The Harrison-Kanji test,
which is an analog of two-factor ANOVA for circular data, was used with FINGER
($4$ levels: $I,M,R,L$) and ANGLE (2 levels, $\alpha_{PE}$ and $\alpha_{PC}$)
as factors. All data analysis were performed in Matlab (MathWorks, Inc.) software.

\subsubsection{The results - acting along the most nimble direction}

During the force--pulse trials, forces of all four fingers changed in parallel
(Figure \ref{FIG9}A). There was a larger change in the force produced by the
instructed finger and smaller changes in the other finger forces. These
patterns are typical of enslaving reported in earlier studies \cite{ZatsiorskyVM(2000),DanionF}.
 PCA applied to the finger force changes
produced similar results over the phase of force increase and the phase of
force drop. The first PC accounted for over $95\%$ of the total variance in the
finger force space in all subjects and for each finger as the instructed
finger. The loading factors at individual finger forces were of the same sign.

In the perturbation conditions, lifting a finger's force sensor produced a
complex pattern of changes in the forces produced by all four fingers (similar
to the results described in \cite{MartinJR,wilhelm}. Typically,
the force of the perturbed finger increased, while the forces produced by the
three other fingers dropped (Figure \ref{FIG9}B). The total force increased.
The first PC accounted for over $95\%$ of the total variance in the finger
force space in all subjects and for each finger as the perturbed finger. The
loading factors at different fingers were of different signs; most commonly,
the perturbed finger loading was of a different sign as compared to the
loading of the three other fingers.

\begin{figure}[ht]
\begin{center}
\includegraphics[width=3in]{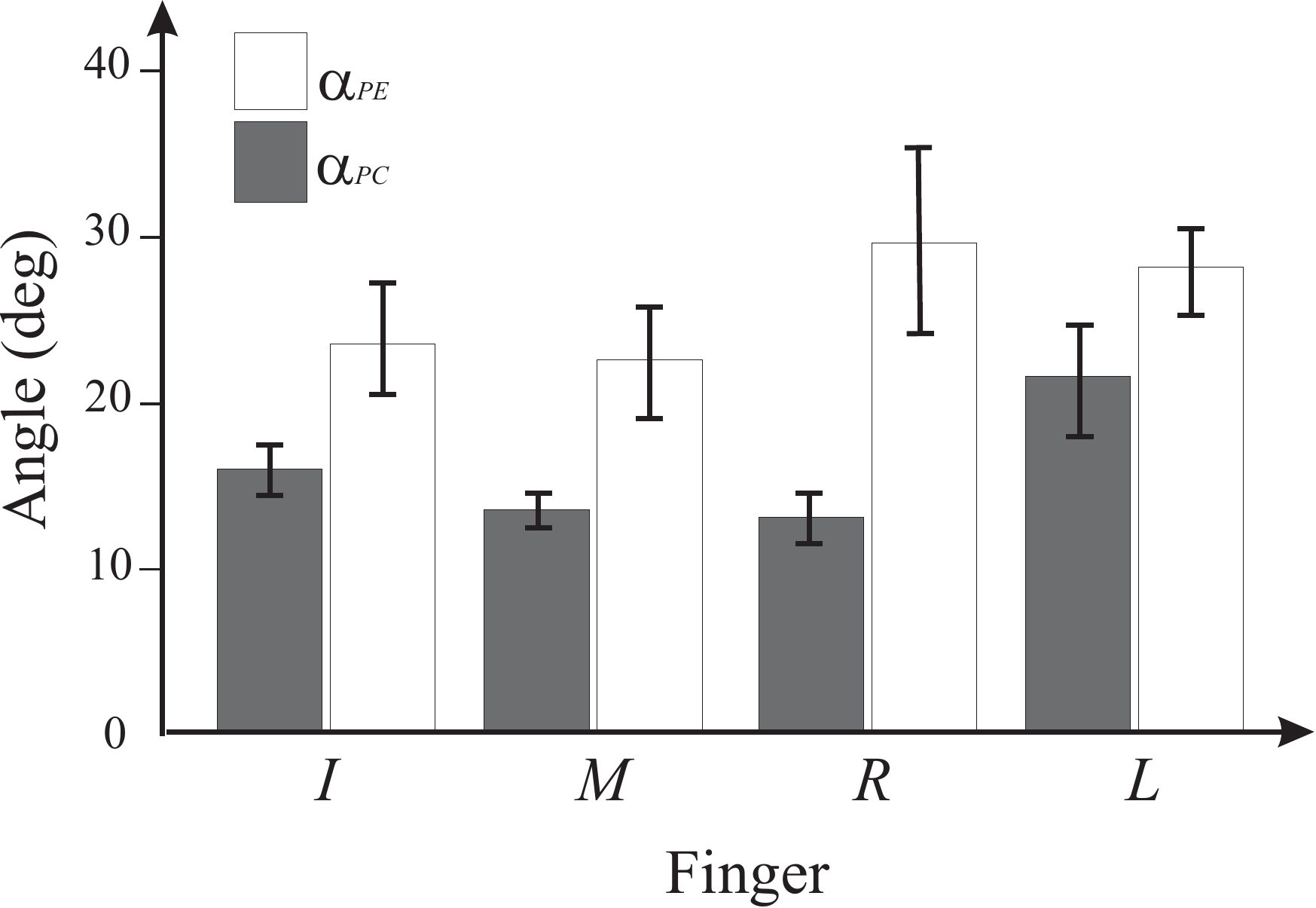}
\caption{The angles between the force vector produced by a quick perturbation
applied to a finger and the force vector during the downward phase of the
force-pulse trial by the same finger ($\alpha_{PE}$) and between the first
vector and the vector of the corrective action ($\alpha_{PC}$). Averaged
across subjects data are shown with standard error bars. Note that for each
finger as the target finger $\alpha_{PE}\;<\;\alpha_{PC}$.}
\label{FIG10}
\end{center}
\end{figure}

Overall, the angle between the vectors of perturbation and correction
($\alpha_{PC}$) was consistently lower than the one between the vectors of
perturbation and voluntary force drop ($\alpha_{PE}$). These results are
illustrated in Figure\ref{FIG10}, which shows averaged (across subjects)
values of the two angles with standard error bars. The gross average of
$\alpha_{PC}$ was $15.9\pm6.6{{}^\circ}$, while it was $25.9\pm10.9{{}^\circ}$
for $\alpha_{PE}$. The Harrison-Kanji test confirmed the main effect of
angle $(F_{[1,56]}=19.17,p<0.0001)$ without other effects.

Overall, these results confirm one of the predictions of the AMN-rule. Indeed,
the first reactions to perturbations in the four-dimensional finger force
space showed relatively small angles with the vector reflecting the effects of
the perturbation on finger forces. This angle was significantly smaller as
compared to typical subject-specific finger force vectors produced when the
subjects tried to increase or decrease force with one finger (the perturbed
finger) only.

\section{Discussion: The Stability Search Algorithm Hypotheses within the
Context of Motor Control Hypotheses}

The most important axiom forming the foundation of our approach is that we
assume the existence of task-specific coordinate systems organized to allow
effective local control. A particular implementation of local control has been
addressed as the \textquotedblleft act on the most nimble\textquotedblright
(AMN) rule. We have shown that this method can solve problems better than
control with random matrices but loses efficacy with an increase in the task
dimensionality, not so much with the system dimensionality. This problem can
be overcome by using feedback with adjustable gain, but in this case the
system slows down dramatically.

Further, we considered a number of additional rules that improve the outcome.
One of them is: If local control does not work, change the coordinate system.
More specific rules that all improve probability of reaching stability include
the following. Deal with one dimension at a time and do not return back to any
of the previously involved dimensions. Organize elemental variables into
generations (groups assembled by links to a specific task variable, $y$) by
their nimbleness. Allowing bi-local control improves the performance even
more. Bi-local control is an example of implementing the general rule of
minimizing non-local actions.

\subsection{Systems of Coordinates in Motor Control}

One of the important features of the suggested scheme is the identification of
three systems of coordinates that can be used to describe processes associated
with the neural control of a movement. Most commonly, movement studies operate
with variables directly measured by the available systems, for example those
that measure kinematic, kinetic, or electromyographic variables. Some of these
variables describe overall performance, for example fingertip coordinates
during pointing. Other variables reflect processes in elements that contribute
to the task-related performance (e.g., joint rotations, digit forces and
moments, muscle activation, etc.). Some variables may not be directly measured
but computed based on other variables and known (or assumed) parameters of the
system (for example, joint torques and muscle forces).

%\vspace{5mm}

Using measured sets of variables to infer coordinate systems used for the
neural control of movement has been a challenge. One of the dominant ideas
originating from the classical studies by Bernstein \cite{BernsteinNA(1967)}
 has been that
elements are united by the central nervous system into relatively stable
groups to reduce the number of parameters manipulated at task-related neural
levels. Such groups have been addressed as \textquotedblleft
synergies\textquotedblright\ \cite{dAvellaA,TingLH(2005)} 
or \textquotedblleft modes\textquotedblright \cite{ZatsiorskyVM(1998),KrishnamoorthyV}.
 Here, we are going to address such groups as
\textquotedblleft modes\textquotedblright\ to avoid confusion with another
definition of synergies as neural structures that ensure stability of
performance based on modes (reviewed in \cite{LatashML(2008)}). Some studies emphasized
the relative invariance of the mode composition across tasks 
\cite{IvanenkoYP,Torres-OviedoG,TingLH(2005)} while other studies showed that the modes
could be rearranged quickly with changes in the stability requirements
\cite{KrishnamoorthyV,DannadosSantosA}. We believe that
some of the disagreements may originate from using the same term for different
coordinate systems.

\vspace{3mm}

Within our scheme, measured variables produced by elements (e.g., digit
forces, joint rotations, and muscle activations) are united into modes that
are relatively stable across task variations. These modes may reflect
preferred changes in the referent body configuration (cf.\cite{FeldmanAG(2009)}) based
on the person's experience with everyday tasks. Mode composition is reflected
in the structure of response to a noisy external input and can be
reconstructed using matrix factorization techniques such as principal
component analysis, factor analysis, independent component analysis, and
non-negative matrix factorization (reviewed in \cite{TingLH(2010)}. Unlike
many earlier studies, we do not assume that the number of modes (the
dimensionality of the space where the control process takes place, $x_{i}$ in
our notation) is smaller than the number of measured variables. It may be
larger. For example, in our experiment, forces of four fingers were measured.
The dimensionality of $x_{i}$ may be higher corresponding, for example, to the
number of muscles or muscle compartments involved in the task.

\vspace{3mm}

According to our main assumption, there is another coordinate system that
allows ensuring stability of performance using local control organized about
each axis. We suggest using a term \textquotedblleft control
coordinates\textquotedblright\ for this system. Unlike modes, control
coordinates are sensitive to task changes, particularly to changes in
conditions that affect stability of performance. When a person encounters a
novel task, however, he/she searches for an adequate set of control
coordinates that would allow implementing local control rules.

\vspace{3mm}

Our experiment showed that a quick reaction to unexpected perturbations acts
along directions in the finger force space that are close to the directions of
finger force deviations produced by the perturbations. In contrast, these
reactions formed larger angles with vectors reflecting finger modes
\cite{DanionF}, eigenvectors in the space of finger forces that reflect finger
force changes when a person tries to act with one finger at a time. This
result corroborates the idea that a quick corrective actions are organized not
along mode directions but along axes of another coordinate system, close to
the ones along which the system shows the quickest deviation in response to
the perturbation.

\subsection{Relations to the Uncontrolled Manifold and Referent Configuration
Hypothesis}

Figure \ref{Fig6t} offers a block diagram related to the control of the hand
based on a few levels. At the upper level, the task is shared between the
actions of the thumb and the opposing fingers represented as a single digit
(virtual finger, Arbib et al. 1985) with the same mechanical effect as the
four fingers combined. Further, the virtual finger action is shared among the
actual fingers (our experiments analyze four-finger coordination at that
level). Even further, action of a finger is shared among a redundant set of
muscles contributing to finger's action. And at the bottom level, each muscle
is organized into a set of motor units by the tonic stretch reflex feedback
that stabilizes the equilibrium point of the system \textquotedblleft muscle +
reflexes + external load\textquotedblright. Only the last level may be seen as
based on relatively well-known neurophysiological mechanisms with the
threshold of the tonic stretch reflex ($\lambda$) serving as a control (input)
variable for the muscle \cite{FeldmanAG(1986)}.

In more intuitive terms, consider controlling the motion of a donkey using a
carrot. The carrot trajectory defines time evolution of referent coordinates
for the head trajectory, to which the head is attracted. But the head cannot
move without moving the legs. So, this big carrot is translated into a
redundant set of smaller mini-carrots for individual legs; and then, into even
more micro-carrots for the joints and muscles. Coordinates of such carrots for
salient body parts represent referent body configuration.

While the scheme in Figure \ref{Fig6t} ensures some stability properties of the action,
changes in the overall organization of action (e.g., changes in the RC
trajectories) may be needed if the task changes or there is a major change in
the external force field. The general principles suggested in this paper offer
a solution for the problem of stabilizing action in such conditions.

A few recent studies have shown that, when a major change in the external
conditions of task execution takes place, corrective actions are seen in both
range (ORT) and self-motion (UCM) spaces with respect to task-specific
performance variables \cite{MattosD(2011),MattosD(2013)}. Moreover, self-motion
dominates, which, by definition, is unable to correct the action. These
observations suggest that no single economy principle can form the foundation
for such corrections. They allow interpretation within the set of principles
suggested in this paper. Any perturbation is expected to induce large effects
in less stable directions (those that span the UCM) as compared to more stable
directions (ORT). According to the AMN-rule, corrective action is organized
along the most nimble of the coordinates that allow local control. Since the
projection of the \textquotedblleft most nimble\textquotedblright\ coordinate
to the UCM is expected to dominate, one can expect the corrective action to be
primarily directed along the UCM as well.

\subsection{Reasonably Sloppy Control May be Good Enough}

Several recent publications presented arguments in favor of the general idea
that the CNS may not solve typical problems perfectly but rather use a set of
simple rules that lead to acceptable solutions for most 
problems\cite{LatashML(2008),LoebGE}. Sometimes, the rules fail to solve specific problems and then
healthy people make mistakes, fall, mishandle objects, spill coffee, etc. We
presented a particular instantiation of such a set of rules (based on the
AMN-rule) and showed that these rules were able to stabilize action with high
probability. The experimental demonstration of relatively small angles between
the vectors of perturbation-induced force changes and corrective changes in
finger forces support the feasibility of the AMN-rule.

Experimental studies of the effects of practice on stability of redundant
systems have shown the existence of two stages (reviewed in \cite{LatashML(2008)}).
First, when a person learns a new task, stability in relevant directions is
developed reflected in an increase in the relative amount of variance along
directions that span the UCM for the salient performance variables. Then, when
accuracy of performance reaches a certain ceiling, further practice leads to a
drop of variance in those seemingly irrelevant directions. Why would a person
stabilize directions that have no clear effect on overall task performance ? We
addressed this issue in the section on perfectionism.

Selection of a particular point (range) within the solution space has been
discussed as resulting from optimizing the action with respect to some
objective (cost) function \cite{PrilutskyBI}. Note that only one
point on the solution hyper-surface is optimal with respect to any given cost
function. Other points within the UCM violate the optimality principle even
though they lead to seemingly perfect performance. In a sense, large variance
within the UCM combined with low variance orthogonal to the UCM implies that
the person is accurate but sloppy. In the course of practice, when the person
is as accurate as one can possibly be with respect to the explicit task,
further practice may stabilize directions within the UCM to ensure that
performance remains as close as possible to optimal with respect to a selected
criterion. This is what we call \textquotedblleft
perfectionism\textquotedblright. Note that perfectionism is never absolute
(\textquotedblleft nobody is perfect\textquotedblright), and the system
remains sloppy, but the degree of sloppiness can be reduced.

We are grateful to Yen-Hsun Wu and Sasha Reschechtko for their help with the experiments.


\begin{thebibliography}{99}                                                                                               

\bibitem {Akulin2014}Akulin VM, (2014) Dynamics of Complex Quantum Systems,
Second Edition, Springer Dordrecht Heidelberg, New York, London, pp. 195-260..

\bibitem {Arbib}Arbib MA, Iberall T, Lyons D (1985) Coordinated control
programs for movements of the hand. In: Goodwin AW and Darian-Smith I, eds.
Hand Function and the Neocortex. Berlin: Springer Verlag; pp. 111-129.

\bibitem {Asaka}Asaka T, Wang Y, Fukushima J, Latash ML (2008) Learning
effects on muscle modes and multi-mode synergies. Experimental Brain Research
184: 323-338.

\bibitem {BernsteinNA1930}Bernstein NA (1930) A new method of mirror
cyclographie and its application towards the study of labor movements during
work on a workbench. Hygiene, Safety and Pathology of Labor, \# 5, p. 3-9, and
\# 6, p. 3-11. (in Russian).

\bibitem {BernsteinNA(1967)}Bernstein NA (1967) The Co-ordination and
Regulation of Movements. Pergamon Press, Oxford

\bibitem {DanionF}Danion F, Sch\"{o}ner G, Latash ML, Li S, Scholz JP,
Zatsiorsky VM (2003) A force mode hypothesis for finger interaction during
multi-finger force production tasks. Biological Cybernetics 88: 91-98.

\bibitem {DannadosSantosA}Danna-Dos-Santos A, Slomka K, Zatsiorsky VM,
Latash ML (2007) Muscle modes and synergies during voluntary body sway.
Experimental Brain Research 179: 533-550.

\bibitem {dAvellaA}d'Avella A, Saltiel P, Bizzi E (2003) Combinations of
muscle synergies in the construction of a natural motor behavior. Nature
Neuroscience 6: 300-308.

\bibitem {FeldmanAG(1986)}Feldman AG (1986) Once more on the
equilibrium-point hypothesis ($\lambda$-model) for motor control. Journal of
Motor Behavior 18: 17-54.

\bibitem {FeldmanAG(2009)}Feldman AG (2009) Origin and advances of the
equilibrium-point hypothesis. Advances in Experimental Medicine and Biology
629: 637-643.

\bibitem {GelfandIM}Gelfand IM, Latash ML (1998) On the problem of adequate
language in movement science. Motor Control 2: 306-313.

\bibitem {IvanenkoYP}Ivanenko YP, Cappellini G, Dominici N, Poppele RE,
Lacquaniti F (2005) Coordination of locomotion with voluntary movements in
humans. Journal of Neuroscience 25: 7238-7253.

\bibitem {KilbreathSL}Kilbreath SL, Gandevia SC (1994) Limited independent
flexion of the thumb and fingers in human subjects. Journal of Physiology 479: 487-497.

\bibitem {KrishnamoorthyV}Krishnamoorthy V, Goodman SR, Latash ML, Zatsiorsky
VM (2003) Muscle synergies during shifts of the center of pressure by standing
persons: Identification of muscle modes. Biological Cybernetics 89: 152-161.

\bibitem {LatashML(2008)}Latash ML (2008) Synergy. Oxford University Press:
New York.

\bibitem {LatashML(2010)}Latash ML (2010) Motor synergies and the
equilibrium-point hypothesis. Motor Control 14: 294-322.

\bibitem {LatashML(2012)}Latash ML (2012) The bliss (not the problem) of
motor abundance (not redundancy). Experimental Brain Research 217: 1-5.

\bibitem {LatashML}Latash ML, Scholz JP, Sch\"{o}ner G (2007) Toward a new
theory of motor synergies. Motor Control 11: 276-308.

\bibitem {LiZM}Li ZM, Latash, ML, Zatsiorsky VM (1998) Force sharing among
fingers as a model of the redundancy problem. Experimental Brain Research 119:276-286

\bibitem {LoebGE}Loeb GE (2012) Optimal isn't good enough. Biological
Cybernetics 106: 757-765.

\bibitem {MartinJR}Martin JR, Budgeon MK, Zatsiorsky VM, Latash ML (2011)
Stabilization of the total force in multi-finger pressing tasks studied with
the `inverse piano' technique. Human Movement Science 30: 446-458.

\bibitem {MartinV}Martin V, Scholz JP, Sch\"{o}ner G (2009) Redundancy,
self-motion, and motor control. Neural Computing 21, 1371-1414.

\bibitem {MattosD(2011)}Mattos D, Latash ML, Park E, Kuhl J, Scholz JP (2011)
Unpredictable elbow joint perturbation during reaching results in multijoint
motor equivalence. Journal of Neurophysiology 106: 1424-1436.

\bibitem {MattosD(2013)}Mattos D, Kuhl J, Scholz JP, Latash ML (2013) Motor
equivalence (ME) during reaching: Is ME observable at the muscle level ? Motor
Control 17:145-175.

\bibitem {PrilutskyBI}Prilutsky BI, Zatsiorsky VM (2002) Optimization-based
models of muscle coordination. Exercise and Sport Science Reviews 30: 32-38

\bibitem {SchieberMH}Schieber MH, Santello M (2004) Hand function: peripheral
and central constraints on performance. Journal of Applied Physiology 96: 2293-300.

\bibitem {ScholzJP}Scholz JP, Sch\"{o}ner G (1999). The uncontrolled manifold
concept: Identifying control variables for a functional task. Experimental
Brain Research 126, 289-306.

\bibitem {SchönerG}Sch\"{o}ner G (1995) Recent developments and problems in
human movement science and their conceptual implications. Ecological
Psychology 8: 291-314.


%added citeREF  ref-1  ref-12  ref-2   ref-22
\bibitem {citeREF}Simon  Models of bounded rationality. MIT Press; Cambridge, MA: 1982.
Ben-Haim Y. Info--gap Decision Theory: Decisions under severe uncertainty. Academic Press; London: 2006.

\bibitem{REF-1} Georgopoulos AP, Kalaska JF, Caminiti R, Massey JT (1982) 
On the relations between the direction of two-dimensional arm movements 
and cell discharge in primate motor cortex. Journal of Neuroscience 2: 1527-1537.

\bibitem{ref-12} Georgopoulos AP, Schwartz AB, Kettner RE (1986) Neural 
population coding of movement direction. 
Science 233: 1416-1419.

\bibitem{REF-2} Schwartz AB, Moran DW (2000) Arm trajectory and representation 
of movement processing in motor cortical activity. Eur J Neurosci. 12(6):1851-6.

\bibitem{ref-22} Amirikian B, Georgopoulos AP (2003) Motor cortex: Coding and decoding of 
directional operations. 
In: Arbib MA (Ed.) The Handbook of Brain Theory and Neural Networks, 
Second Edition, pp. 690-701, MIT Press: Cambridge, MA.



\bibitem {SlobounovS}Slobounov S, Chiang H, Johnston J, Ray W (2002)
Modulated cortical control of individual fingers in experienced musicians: an
EEG study. Electroencephalographic study. Clinical Neurophysiology 113: 2013-2024.

\bibitem {Stain}Stein RB, Gossen ER, Jones KE (2005) Neuronal variability:
noise or part of the signal? Nature Reviews in Neuroscience 6(5): 389-97.

\bibitem {TerekhovAV}Terekhov AV, Pesin YB, Niu X, Latash ML, Zatsiorsky VM
(2010) An analytical approach to the problem of inverse optimization: An
application to human prehension. Journal of Mathematical Biology 61: 423-453.

\bibitem {TingLH(2010)}Ting LH, Chvatal SA (2010) Decomposing muscle
activity in motor tasks: Methods and Interpretation. In: Danion F, Latash ML
(Eds) Motor Control: Theories, Experiments, and Applications, p. 102-139,
Oxford University Press: New York, NY.

\bibitem {TingLH(2005)}Ting LH, Macpherson JM (2005) A limited set of muscle
synergies for force control during a postural task. Journal of Neurophysiology
93: 609-613.

\bibitem {Torres-OviedoG}Torres-Oviedo G, Ting LH (2007) Muscle synergies
characterizing human postural responses. Journal of Neurophysiology 98: 2144-2156

\bibitem{wilhelm} Wilhelm L., Zatsiorsky V.M., Latash M.L. (2013) Equifinality and its violations in a redundant system: Multi-finger accurate force production. Journal of Neurophysiology 110: 1965-1973.

\bibitem {WuYH}Wu YH, Pazin N, Zatsiorsky VM, Latash ML (2013) Improving
finger coordination in young and elderly persons. Experimental Brain Research
226: 273--283.

\bibitem {ZatsiorskyVM(1998)}Zatsiorsky VM, Li ZM, Latash ML (1998)
Coordinated force production in multi-finger tasks: Finger interaction and
neural network modeling. Biological Cybernetics 79: 139-150.

\bibitem {ZatsiorskyVM(1999)}Zatsiorsky VM, Duarte M (1999) Instant
equilibrium point and its migration in standing tasks: rambling and trembling
components of the stabilogram. Motor Control. 3(1):28-38.

\bibitem {ZatsiorskyVM(2000)}Zatsiorsky VM, Li ZM, Latash ML (2000).
Enslaving effects in multi-finger force production. Experimental Brain
Research 131: 187-195.
\end{thebibliography}
\end{document}